\documentclass[prb,twocolumn,floatfix,longbibliography]{revtex4-2}
\usepackage[english]{babel}
\usepackage[utf8]{inputenc}
\usepackage[T1]{fontenc}
\usepackage{amsmath}
\usepackage{amssymb}
\usepackage{graphicx}
\usepackage{bm}
\usepackage{color}
\usepackage{hyperref}
\usepackage{upgreek}
\usepackage{scalerel}
\usepackage{pgffor}
\usepackage{pdfpages}
\usepackage{float}
\usepackage{appendix}
\usepackage{physics} 
\usepackage{lscape}   
\usepackage{silence}
\usepackage{soul,xcolor} %strikethrough
\setstcolor{red} %strikethrough color
\usepackage{comment}  
\makeatletter
\AtBeginDocument{\let\LS@rot\@undefined}
\makeatother

\begin{document}
\title{Exceptional degeneracies in non-Hermitian Rashba semiconductors}
 \author{Jorge Cayao} 
 \email[ ]{ jorge.cayao@physics.uu.se}
\affiliation{Department of Physics and Astronomy, Uppsala University, Box 516, S-751 20 Uppsala, Sweden}

\date{\today}
\begin{abstract}
Exceptional points are spectral degeneracies of non-Hermitian systems  where eigenvalues and eigenvectors coalesce, inducing  unique topological phases that have no counterpart in the Hermitian realm. Here we consider a non-Hermitian system by coupling a two-dimensional semiconductor with Rashba spin-orbit coupling to a ferromagnet lead  and show the emergence of highly tunable   exceptional points along rings in momentum space. Interestingly, these exceptional degeneracies are the endpoints of lines formed by the eigenvalue coalescence at finite real energy, resembling  the bulk Fermi arcs commonly defined at zero real energy. We then show that an in-plane Zeeman field provides a  way to control these exceptional degeneracies although higher values of non-Hermiticity are required in contrast to the zero Zeeman field regime. Furthermore,  we find that the  spin projections also coalescence  at  the exceptional degeneracies and can acquire larger values than in the Hermitian regime.  Finally, we demonstrate that the  exceptional degeneracies induce large spectral weights, which can be used as a signature for their detection. Our results thus reveal the potential of systems with Rashba spin-orbit coupling for realizing non-Hermitian bulk phenomena.
\end{abstract}\maketitle

%%%%%%%%%%%%%%%%%%%%%%%%%%%%%%%
%SECTION I:                  INTRODUCTION                            %
%%%%%%%%%%%%%%%%%%%%%%%%%%%%%%%
\section{Introduction}
\label{section:Intro}
The effect of dissipation, often seen as detrimental, has recently attracted a paramount attention in physics due its potential to induce novel  phenomena with technological applications  \cite{el2018non,ozdemir2019parity,RevModPhys.93.015005,doi:10.1080/00018732.2021.1876991,parto2020non,wiersig2020review,ding2022non}. Dissipation naturally occurs in open systems and is effectively described by non-Hermitian (NH) Hamiltonians \cite{PhysRevX.8.031079, PhysRevB.99.235112, PhysRevX.9.041015}. The most salient property of these NH models is the emergence of  a complex spectrum with degeneracies known as exceptional points (EPs) \cite{TKato, heiss2004exceptional, berry2004physics, Heiss_2012, PhysRevLett.86.787, PhysRevLett.103.134101, PhysRevLett.104.153601, gao2015observation, doppler2016dynamically,PhysRevB.99.121101,arouca2022exceptionally}, where eigenstates and eigenvalues coalesce, in stark contrast to Hermitian systems.  While EPs were initially seen as a mathematical curiosity, it has been recently shown that they represent truly topological objects enabling  topological phases with no counterpart in  Hermitian setups \cite{RevModPhys.93.015005,doi:10.1080/00018732.2021.1876991,ding2022non}. 

The concept of EPs and their topological properties have recently been generalized to  higher dimensions, giving rise to exceptional degeneracies in the form of lines, rings, and surfaces as generic and stable bulk phenomena. These exceptional degeneracies have already proven crucial to enable unique topological effects
 \cite{RevModPhys.93.015005,doi:10.1080/00018732.2021.1876991}, such as   enhanced sensing  \cite{hodaei2017enhanced,chen2017exceptional}, unidirectional lasing \cite{peng2016chiral,Longhi:17}, and bulk Fermi arcs \cite{kozii2017non,PhysRevB.97.014512,PhysRevB.98.035141,PhysRevB.99.041202,PhysRevLett.123.066405,PhysRevLett.123.123601,science359Zhou,doi:10.7566/JPSCP.30.011098, PhysRevLett.125.227204, PhysRevLett.127.186601,PhysRevLett.127.186602,yoshida2020exceptional,rausch2021exceptional}, which do not have a Hermitian analog. Despite the numerous theoretical and experimental studies, however, the majority of them has  investigated exceptional degeneracies mostly in optical and photonic systems \cite{el2018non,ozdemir2019parity,wiersig2020review,parto2020non}.  
 
Material junctions have  been shown to offer another  powerful and experimentally relevant platform for the realization of exceptional degeneracies \cite{pikulin2012topological, pikulin2013two, san2016majorana, avila2019non, PhysRevB.98.155430, PhysRevB.98.245130, PhysRevResearch.1.012003,PhysRevB.105.094502,cayao2022NH}. Material junctions constitute electronic open systems with a clear NH description that is well-established in quantum transport \cite{datta1997electronic}.  In this regard, open semiconductor-superconductor junctions have  been shown to host several classes of exceptional degeneracies \cite{san2016majorana, avila2019non,cayao2022NH}, which characterize distinct NH topological phases without analog in the Hermitian regime.  Notwithstanding the importance of this study,  it  only focused on the impact of non-Hermiticity on the  superconducting properties, such as on its particle-hole symmetry and energy gap, leaving largely unexplored the role of non-Hermiticity on the semiconductor.  Of particular importance in such  semiconductors is their intrinsic Rashba spin-orbit coupling (SOC) \cite{PhysRev.100.580,Rashba1960,rashba84a}, which arises due to the lack of structural inversion symmetry and induces a  spin-momentum locking \cite{winkler2003spin,chen2021spin}.  This  property of the Rashba SOC has been shown to enable a great  control  of the electron's spin, a crucial ingredient for several spintronics and topological phenomena \cite{manchon2015new}, already proven useful in recent experiments \cite{manchon2015new,Lutchyn2018Majorana, prada2019andreev, flensberg2021engineered,PhysRevMaterials.2.044202,chen2021spin}. However, despite the advances,   the interplay between  Rashba SOC and non-Hermiticity still remains unknown, specially the potential of this combination for inducing exceptional degeneracies.

%%%%%%%%%%%%%%%%%
%Fig 1
%%%%%%%%%%%%%%%%%%
\begin{figure}[!t]
	\centering
	\includegraphics[width=0.3\textwidth]{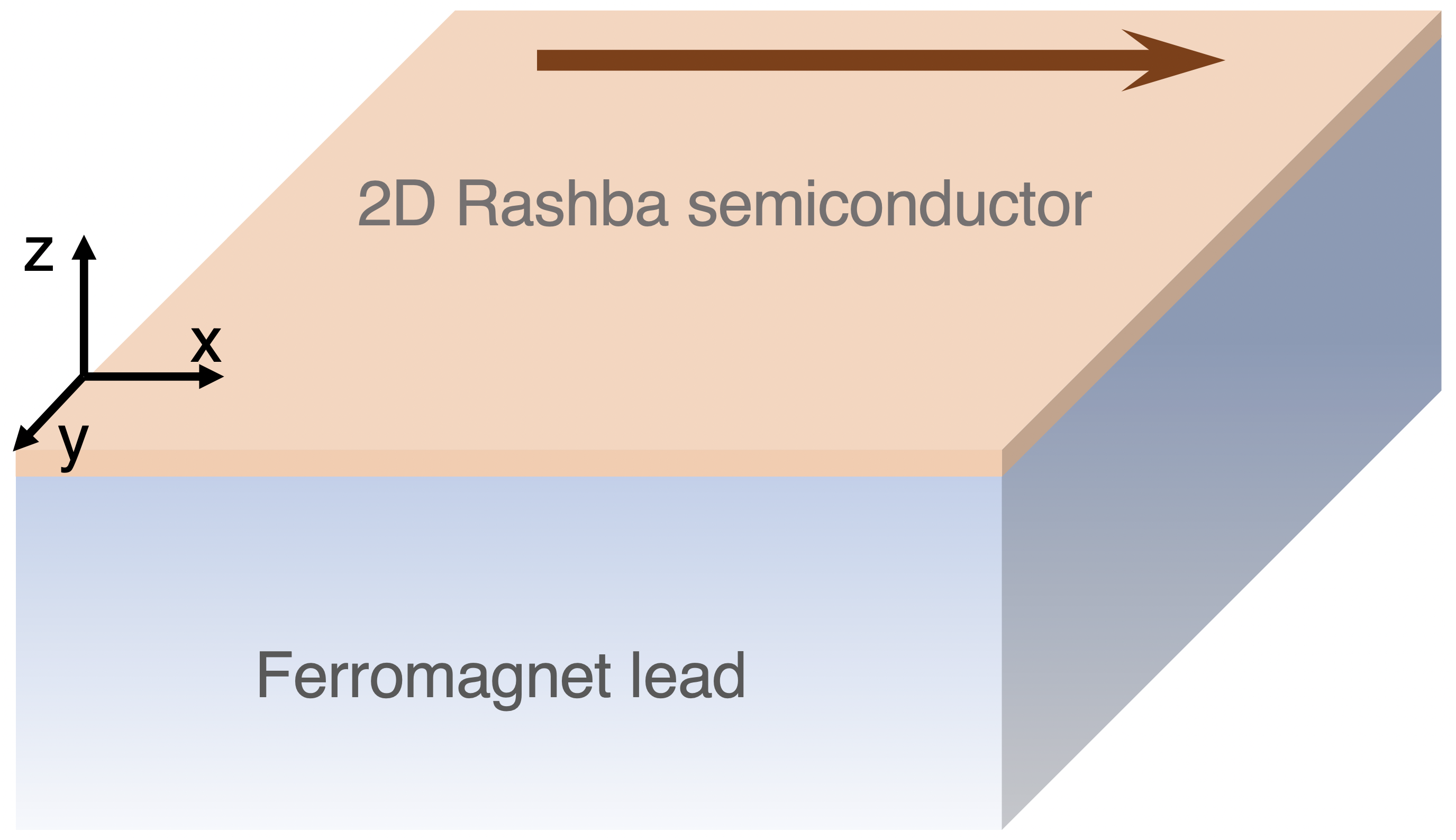} 
	\caption{Schematics of studied non-Hermitian Rashba system: a 2D Rashba semiconductor (orange) is coupled to a semi-infinite ferromagnet lead (gray). A Zeeman field along $x$ is applied (brown) in order to control the emergent non-Hermitian degeneracies.}
	\label{fig1}
\end{figure}

In this work we consider a realistic NH  Rashba semiconductor  and discover the formation of stable and highly tunable bulk exceptional degeneracies. In particular, we engineer a NH  Rashba system  by coupling a two-dimensional (2D) semiconductor  with Rashba SOC to a semi-infinite ferromagnet lead, an easily achievable heterostructure using e.g., InAs or InSb semiconductors \cite{manchon2015new,Lutchyn2018Majorana, prada2019andreev, flensberg2021engineered,PhysRevMaterials.2.044202,chen2021spin}, see also \cite{kammhuber2017conductance}. We discover that EPs appear along rings in 2D momentum space and mark the ends of lines formed by the coalescence of eigenvalues at  finite real energy. 
The emergence of eigenvalues at the same  real energy resembles the formation of bulk Fermi arcs, which, although initially conceived at zero real energy,  have recently been generalized   to finite real energies \cite{PhysRevResearch.4.L022064}. We also show that the exceptional degeneracies found here can be controlled by an in-plane Zeeman field but then higher values of non-Hermiticity are required.  Furthermore, we  find that the spin projections coalesce at the exceptional degeneracies and can even develop larger values than in the Hermitian phase due to non-Hermiticity.  Finally, we find that the exceptional degeneracies induce large spectral features, which can be detected, e.g., using angle-resolved photoemission spectroscopy (ARPES).

%%%%%%%%%%%%%%%%%%%%%%%%%%%%%%%%%%%%
% SECTION II:                        MODEL AND METHODS                       %
%%%%%%%%%%%%%%%%%%%%%%%%%%%%%%%%%%%%
\section{Non-Hermitian effective model}
\label{sec:Model}
We consider an open system by coupling a 2D semiconductor with Rashba SOC to a semi-infinite ferromagnet lead, as schematically shown in Fig.\,\ref{fig1}. This open system  is modelled by the following effective NH Hamiltonian
\begin{equation}
\label{openNH}
H_{\rm eff}=H_{\rm R}+\Sigma^{r}(\omega=0)\,,
\end{equation}
where $H_{\rm R}$ describes the closed  system, which is Hermitian, and $\Sigma^{r}(\omega=0)$ is the zero-frequency retarded self-energy due to the coupling to the semi-infinite ferromagnet lead. More specifically, the closed system corresponds to a 2D Rashba semiconductor described by
\begin{equation}
\label{Hclosed}
H_{\rm R}=\xi_{k}+\alpha(k_{y}\sigma_{x}-k_{x}\sigma_{y})\,,
\end{equation}
where $\xi_{k}=\hbar^{2}(k_{x}^{2}+k_{y}^{2})/2m-\mu$ is the kinetic energy, $k_{x(y)}$ the momentum along $x(y)$, $\mu$ is the chemical potential, $\alpha$ is the Rashba SOC strength, and $\sigma_{j}$ the $j$-th  Pauli matrix in spin space, and without loss of generality we assume $\hbar=m=1$. The Hamiltonian $H_{\rm R}$ in Eq.\,(\ref{Hclosed}) describes well the Rashba SOC in 2D semiconductors, such as in InAs or InSb, which are also within   experimental reach   \cite{manchon2015new,Lutchyn2018Majorana, prada2019andreev, flensberg2021engineered,PhysRevMaterials.2.044202,chen2021spin,kammhuber2017conductance}.
As an external control knob, we  also consider that the closed system is  subjected to an applied magnetic field along $x$ which produces a  Zeeman field $B$, denoted by the  brown arrow in Fig.\,\ref{fig1}. The effect of this Zeeman field is modelled by adding $B\sigma_{x}$ to $H_{\rm R}$ in Eq.\,(\ref{Hclosed}) which induces a renormalization to the SOC term $\alpha k_{y}\sigma_{x}$.

The zero-frequency self-energy $\Sigma^{r}(\omega=0)$ in Eq.\,(\ref{openNH}), whose independence of frequency $\omega$ is well justified in the wide-band limit \cite{datta1997electronic}, is analytically obtained and given by \cite{PhysRevResearch.1.012003, PhysRevB.105.094502}
\begin{equation}
\label{NHSelf}
\Sigma^{r}(\omega=0)=-i \Gamma \sigma_{0}-i\gamma \sigma_{z}\,,
\end{equation}
where $\Gamma=(\Gamma_{\uparrow}+\Gamma_{\downarrow})/2$ and $\gamma=(\Gamma_{\uparrow}-\Gamma_{\downarrow})/2$, with $\Gamma_{\sigma}=\pi|t'|^{2}\rho_{\rm L}^{\sigma}$, being $t'$ the hopping amplitude into the lead from the 2D Rashba semiconductor and $\rho_{\rm L}^{\sigma}$  the surface  density of states of the lead for spin $\sigma=\uparrow, \downarrow$. It is thus evident that $\Gamma_{\sigma}$ characterizes the coupling amplitude between the lead and the 2D Rashba semiconductor.  For completeness, the derivation of the self-energy given by Eq.\,(\ref{NHSelf}) is presented in Appendix \ref{Appendix}.

The self-energy in Eq.\,(\ref{NHSelf}) is imaginary and thus NH, a unique effect emerging due to the coupling to the semi-infinite ferromagnet lead.  Thus, the imaginary self-energy renders the total effective Hamiltonian $H_{\rm eff}$ to be NH, introducing dramatic changes in the properties of the closed system $H_{\rm R}$, which is the focus of this work here.  In particular, we are interested in investigating the interplay between non-Hermiticity and Rashba SOC in 2D semiconductors and how it leads to the formation of  bulk exceptional  degeneracies.

%%%%%%%%%%%%%%%%%
%Fig 2
%%%%%%%%%%%%%%%%%%
\begin{figure}[!t]
	\centering
	\includegraphics[width=0.495\textwidth]{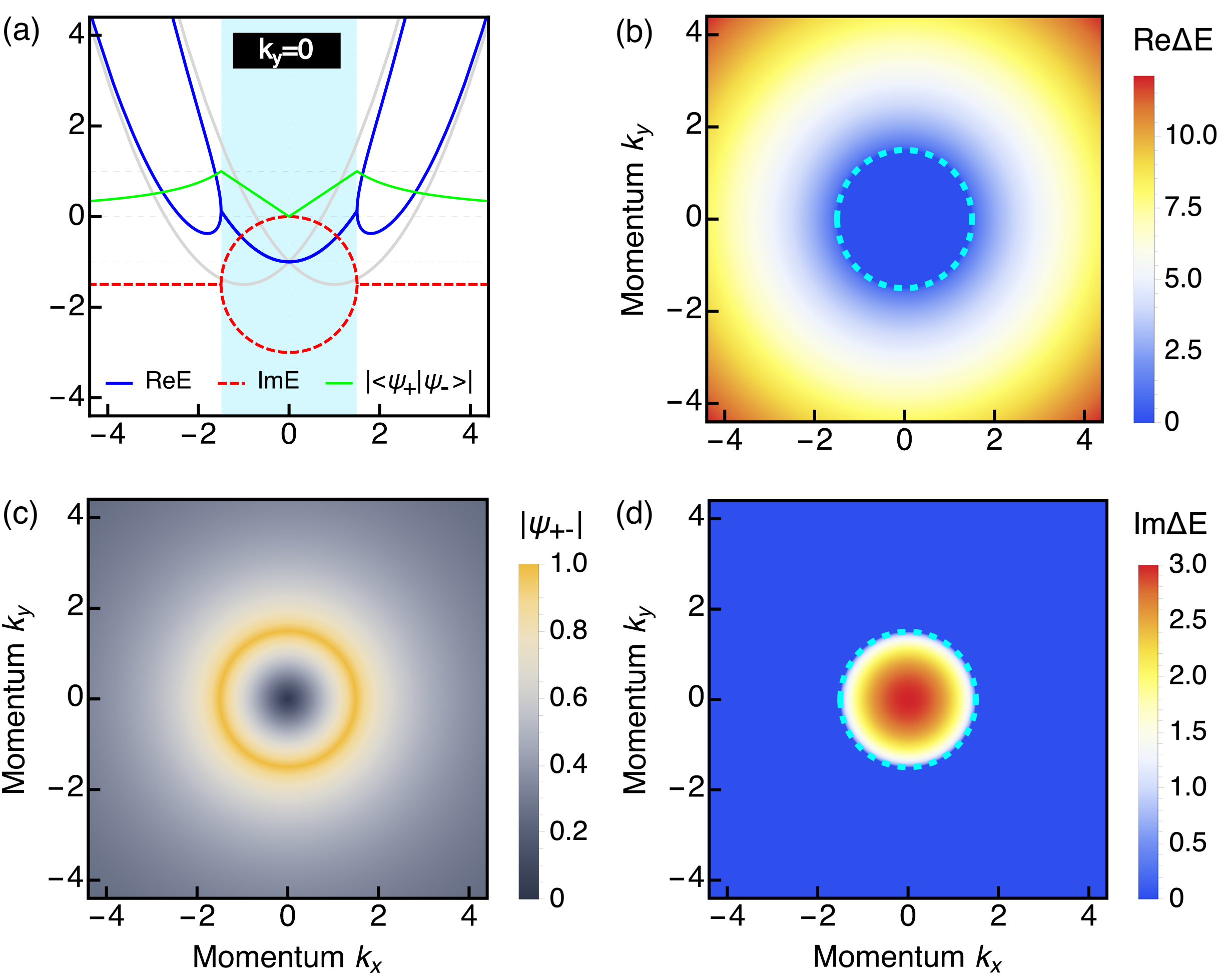} 
	\caption{Exceptional degeneracies in 2D Rashba semiconductors at zero Zeeman field: (a) Real (Re) and imaginary (Im) parts of the eigenvalues as a function of $k_{x}$ depicted in solid blue and dashed red curves at $k_{y}=0$. Green curve represents the absolute value of the overlap between the two wavefunctions $\psi_{+-}=\bra{\psi_{+}}\ket{\psi_{-}}$. Gray curves show eigenvalues without non-Hermiticity, $\Gamma_{\uparrow/\downarrow}=0$. (b,d) Real and imaginary parts of the energy differences $\Delta E=(E_{+}-E_{-})$ as a function of $k_{x}$ and $k_{y}$.  (c) represents $\psi_{+-}$ as a function of $k_{x}$ and $k_{y}$. Parameters: $\alpha=1$, $\Gamma_{\uparrow}=3$, $\Gamma_{\downarrow}=0$, $\mu=1$, $B=0$.}
	\label{fig2}
\end{figure}

%%%%%%%%%%%%%%%%%%%%%%%%%%%%%%%%%%%%
% SECTION III:           EXCEPTIONAL DEGENERACIES                     %
%%%%%%%%%%%%%%%%%%%%%%%%%%%%%%%%%%%%
\section{Exceptional degeneracies}
\label{sec:sectionIII}
To identify the emergence of bulk exceptional degeneracies,  we obtain the eigenvalues and eigenvectors of the effective Hamiltonian $H_{\rm eff}$ given by Eq.\,(\ref{openNH}).  At zero Zeeman field $B=0$, they are given by
\begin{equation}
\label{EnergiesNH}
E_{\pm}=\xi_{k}-i\Gamma\pm\sqrt{\alpha^{2}|k|^{2}-\gamma^{2}}\,,
\end{equation}
\begin{equation}
\label{EvectorsNH}
\Psi_{\pm}=\frac{1}{\sqrt{2}}
\begin{pmatrix}
1\\
\frac{i\gamma\pm\sqrt{\alpha^{2}|k|^{2}-\gamma^{2}}}{\alpha(k_{y}+ik_{x})}
\end{pmatrix}\,,
\end{equation}
where  $|k|^{2}=k_{x}^{2}+k_{y}^{2}$ and  $\pm$   labels the two distinct bands which have a mixture of $\uparrow$ and $\downarrow$ spins. At finite Zeeman fields $B$, the eigenvalues  and eigenvectors can be obtained by replacing $\alpha k_{y}\rightarrow B+\alpha k_{y}$ in Eqs.\,(\ref{EnergiesNH}) and (\ref{EvectorsNH}).  
An immediate observation in the energies and wavefunctions  is their dependence on the couplings $\Gamma_{\uparrow,\downarrow}$ via $\gamma$ and $\Gamma$, already revealing a clear impact of the NH self energy given by Eq.\,(\ref{NHSelf}).  
This can be visualized in Fig.\,\ref{fig2}, where we plot the eigenvalues and eigenvectors as a function of momenta $k_{x}$ and $k_{y}$  at zero Zeeman field. At $\Gamma_{\uparrow,\downarrow}=0$, the system described by   Eq.\,(\ref{openNH}) is Hermitian and its two eigenvalues in Eq.\,(\ref{EnergiesNH}) are real: they correspond to two parabolas shifted by $\pm k_{\rm soc}=\pm m\alpha/\hbar^{2}$ that intersect at $k_{x,y}=0$, see gray curves in Fig.\,\ref{fig2}(a). Here, their respective eigenvectors  are orthogonal as expected for Hermitian systems, see  Eq.\,(\ref{EvectorsNH}). While this Rashba system is gapless at zero momenta, finite values of $k_{y}$   opens a gap at $k_{x}=0$ even at zero Zeeman field. A finite in-plane Zeeman field opens a gap at $k_{x,y}=0$, also known as helical gap, where states  are counter propagating and have distinct spins \cite{PhysRevB.91.024514,kammhuber2017conductance,oshima2018tunneling,oshima2019unconventional}.

At any $\Gamma_{\uparrow,\downarrow}\neq0$, the two eigenvalues $E_{\pm}$    acquire  finite imaginary parts that strongly depend on momenta, see  Eq.\,(\ref{EnergiesNH}). The formation of eigenvalues with imaginary terms signals the emergence of NH physics as a pure effect due to the ferromagnet lead \cite{PhysRevResearch.1.012003, PhysRevB.105.094502,cayao2022NH}. The inverse of these imaginary parts define the quasiparticle lifetime in the 2D Rashba semiconductor, thus offering a clear physical meaning of non-Hermiticity \cite{datta1997electronic}. From the dependence of the eigenvalues on $\gamma$ in Eq.\,(\ref{EnergiesNH}), we note that their imaginary parts exhibit a non trivial behaviour. In fact, at $\gamma=0$, which is satisfied when $\Gamma_{\uparrow}=\Gamma_{\downarrow}$, the two eigenvalues acquire the same imaginary part equal to $-i\Gamma$. This situation remains for $\gamma\neq0$ only when $|\gamma|<\alpha|k|$. At these conditions, therefore, quasiparticles in the Rashba semiconductor have the same and constant lifetime.  

 %%%%%%%%%%%%%%%%%
%Fig 3
%%%%%%%%%%%%%%%%%%
\begin{figure}[!t]
	\centering
	\includegraphics[width=0.49\textwidth]{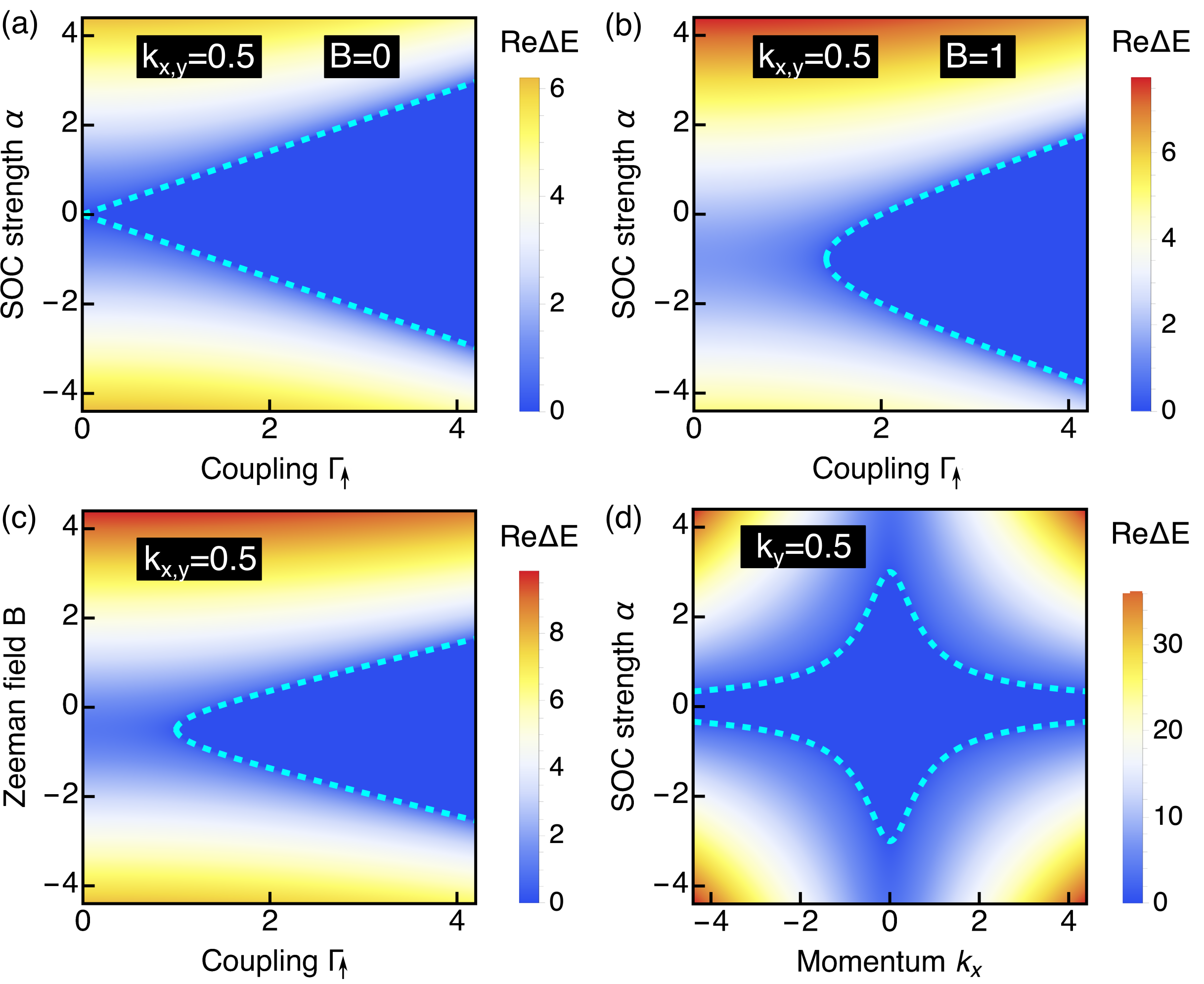} 
	\caption{Tunability of exceptional degeneracies in 2D semiconductors with Rashba SOC: (a,b) Real part of the energy difference $\Delta E=(E_{+}-E_{-})$ as a function of  $\alpha$ and $\Gamma_{\uparrow}$ at $B=0$ and $B=1$. (c,d) Same quantity as in (a,b) but as a function of $B$ and $\Gamma_{\uparrow}$ at $\alpha=1$ (c), and as a function of $\alpha$ and $k_{x}$  at $B=0$ (d). Parameters:  $\Gamma_{\downarrow}=0$, $\mu=1$.}
	\label{fig3}
\end{figure}

 The behaviour of the eigenvalues becomes more interesting when $\gamma\neq0$ and $|\gamma|>\alpha|k|$, which then allows   the two eigenvalues to acquire distinct imaginary parts.  This is visualized in Fig.\,\ref{fig2}(a) where we plot the real and imaginary parts of the eigenvalues at zero Zeeman field and at $k_{y}=0$, see solid blue and dashed red curves. Surprisingly, we observe that both the real and imaginary parts simultaneously merge at finite energy into a single value at special positive and negative momenta.  The regime with $\gamma\neq0$ and $|\gamma|>\alpha|k|$ not only affects the eigenvalues but also the eigenvectors, which can be noticed by inspecting their inner product or overlap $\psi_{+-}=\bra{\psi_{+}}\ket{\psi_{-}}$,  $\psi_{\pm}$ is given by Eq.\,(\ref{EvectorsNH}). This overlap is depicted by the green curve in Fig.\,\ref{fig2}(a), where we see that it reaches $1$ at the special momenta, a situation that can only occur if the eigenvectors are parallel. Although unusual,  the  behaviour of eigenvectors and eigenvalues seen in Fig.\,\ref{fig2}(a)  is common in NH Hermitian systems. In particular, the spectral  degeneracies occurring at the special momenta discussed here  signal the emergence of EPs, whose formation can be understood by noting that they occur when  the square root in Eq.\,(\ref{EnergiesNH}) vanishes. At zero Zeeman field, the  condition for the formation of EPs is given by
 \begin{equation}
\label{EPcondition}
\alpha^{2}(k_{x}^{2}+k_{y}^{2})-\gamma^{2}=0\,,
\end{equation}
while at finite Zeeman field we have to change $\alpha k_{y}\rightarrow B +\alpha k_{y}$. 
 At this EP condition, the eigenvalues and eigenvectors become,
\begin{equation}
\label{EPEvalPhi}
\begin{split}
E^{\rm EP}_{\pm}&=\xi_{k}-i\Gamma\,,\\
 \Psi^{\rm EP}_{\pm}&=\frac{1}{\sqrt{2}}
\begin{pmatrix}
1\\
\frac{i\gamma}{\alpha(k_{y}+ik_{x})}
\end{pmatrix}\,,
\end{split}
\end{equation}
where the values of momenta satisfy the condition given by Eq.\,(\ref{EPcondition}).
Hence, the two eigenvalues (eigenvectors) coalesce at EPs: instead of having two eigenvalues (eigenvectors), only one eigenvalue (eigenvector) remains at EPs, see Fig.\,\ref{fig2}(a). 
 For $k_{y}=0$, the EP occur at positive and negative momenta given by  $\pm k^{\rm EP}_{x}=\pm(\gamma/\alpha)^{2}$ at $B=0$, marking the ends of the cyan region in Fig.\,\ref{fig2}(a). Between these two EP points, the eigenvalues have the same real part determined by the quadratic dispersion $\xi_{k}$ and  different imaginary parts determined by $-i\Gamma\pm i\sqrt{\gamma^{2}-\alpha^{2}|k|^{2}}$, see cyan region in Fig.\,\ref{fig2}(a).  In relation to the eigenvectors, the fact that both  merge into a single eigenvector implies that they are parallel, a situation that has no analog in the Hermitian regime but   expected at EPs of NH systems \cite{RevModPhys.93.015005}. This effect can be also seen in the inner product (or overlap) between the two eigenvectors $\bra{\psi_{+}}\ket{\psi_{-}}$ in Eq.\,(\ref{EvectorsNH}),  where it  reaches 1 at the EPs, see green curve in Fig.\,\ref{fig2}(a). We remark that for EPs to emerge it is crucial to have eigenvalues with different imaginary parts, which is only achieved when $\Gamma_{\uparrow}\neq\Gamma_{\downarrow}$, thus pointing out the necessity of a ferromagnet lead for the NH features discussed in Fig.\,\ref{fig2}(a).    In passing, we note that having bulk energy lines due to eigenvalues with the same real part resembles the formation of bulk Fermi arcs \cite{kozii2017non,PhysRevB.97.014512,PhysRevB.98.035141,PhysRevB.99.041202,PhysRevLett.123.066405,PhysRevLett.123.123601,science359Zhou,doi:10.7566/JPSCP.30.011098, PhysRevLett.125.227204, PhysRevLett.127.186601,PhysRevLett.127.186602}, although here they occur at finite real energy in contrast to the common expectation at zero real energy.  
  In this regard, very recently, the definition of bulk Fermi arcs has been generalized to any two eigenvalues with the same real energy \cite{PhysRevResearch.4.L022064}, suggesting that the bulk energy lines found here might   be an example of bulk Fermi arcs.  However, the detail properties of this NH bulk effect require a throughout investigation  which will be addressed elsewhere.

Furthermore, another property of the EPs determined by the  condition in Eq.\,(\ref{EPcondition}) is that  they occur 
along a ring defined by $\alpha^{2}(k_{x}^{2}+k_{y}^{2})=\gamma^{2}$ at $B=0$ or by $\alpha^{2}k_{x}^{2}+(\alpha k_{y}+B)^{2}=\gamma^{2}$ at $B\neq0$. To support this idea, in Fig.\,\ref{fig2}(b,d) we plot the difference between real and imaginary parts of the eigenvalues, namely, ${\rm Re}\Delta E={\rm Re}(E_{+}-E_{-})$ and ${\rm Im}\Delta E={\rm Im}(E_{+}-E_{-})$, as a function of $k_{x}$ and $k_{y}$. In this case, the blue regions indicate  ${\rm Re}\Delta E=0$ and ${\rm Im}\Delta E=0$, with their borders marking the occurrence of  rings. To highlight these rings, in Fig.\,\ref{fig2}(b,d)   we  also plot the  condition given by Eq.\,(\ref{EPcondition}) in dashed cyan color.  The nature of these rings can be also seen in Fig.\,\ref{fig2}(c), where we   plot the eigenvector overlap, which acquires $1$ exactly along them which implies that the eigenvectors here become parallel  as at the EPs discussed previously. This thus demonstrates that the rings seen in Fig.\,\ref{fig2}   truly  represent bulk exceptional  degeneracies of NH 2D Rashba semiconductors and can be referred to as exceptional rings.

To induce the formation of the exceptional degeneracies, or exceptional rings, it is sufficient  the interplay between non-Hermiticity and SOC, as clearly seen in Eq.\,(\ref{EPcondition}). While this conclusion is already evident in Fig.\,\ref{fig2}, to further support it, in Fig.\,\ref{fig3}(a) we present ${\rm Re}\Delta E$ as a function of the SOC strength $\alpha$ and coupling $\Gamma_{\uparrow}$ at finite momenta and zero Zeeman field. We obtain that  the region with ${\rm Re}\Delta E=0$ increases following a triangular-shaped profile depicted in blue, which is delimited by $\pm \sqrt{\gamma^{2}/|k|^{2}}$ indicated by cyan dashed lines.  At fixed momenta,  the SOC drives the formation of EPs, requiring lower SOC when non-Hermiticity is small. By fixing only one momentum coordinate, it is also possible to induce EPs, as seen in Fig.\,\ref{fig3}(d). Furthermore, another possibility to control the appearance of EPs is by  an in-plane Zeeman field along $x$ as considered in Fig.\,\ref{fig1}. Thus, in Fig.\,\ref{fig3}(b) we show   ${\rm Re}\Delta E$ as a function of $\alpha$ and $\Gamma_{\uparrow}$ at finite  $B$, while  in Fig.\,\ref{fig3}(c) we show ${\rm Re}\Delta E$ as a function of $B$ and $\Gamma_{\uparrow}$. In this case, we identify two relevant features. First, at finite SOC and finite momenta, the non-Hermiticity needed to induce EPs needs to overcome the effect of the Zeeman field $B$, thus requiring larger non-Hermiticity than in the absence of $B$ [Fig.\,\ref{fig3}(b)]. Second, at all  fixed parameters, the Zeeman field  drives the emergence of EPs [Fig.\,\ref{fig3}(b)]; the EPs here are marked by the cyan curves which correspond to $-\alpha k_{y}\pm\sqrt{\gamma^{2}+\alpha^{2}k_{x}^{2}}$. In sum, the bulk exceptional degeneracies found in 2D  Rashba semiconductors exhibit a high degree of tunability  by SOC, momenta, and Zeeman field, which could be  relevant for their realization and subsequent observation.

 \section{Spin projections}
\label{sec:sectionVI}
\begin{figure}[!t]
	\centering
	\includegraphics[width=0.49\textwidth]{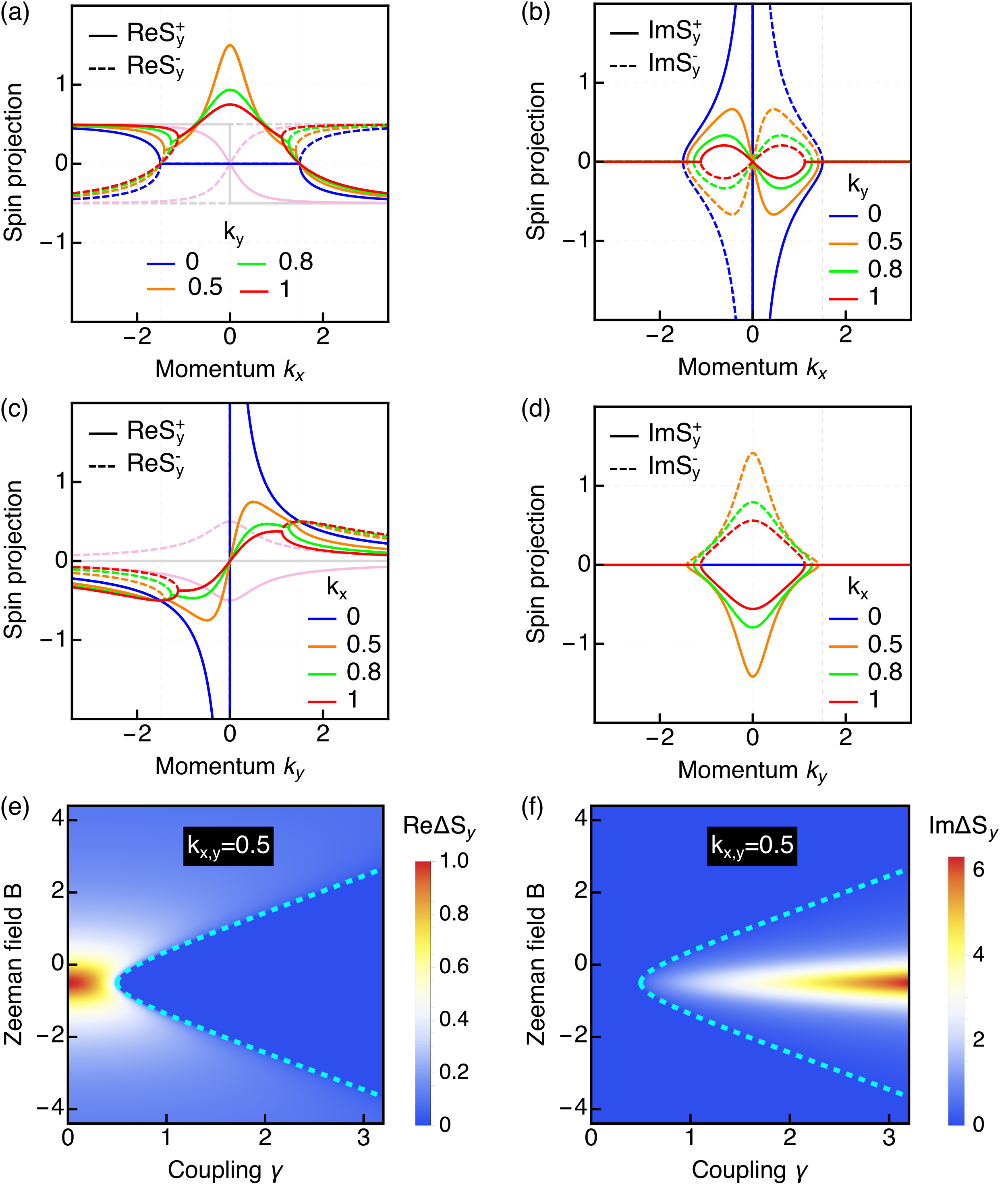} 
	\caption{Spin projection along $y$, $S_{y}^{\pm}$: (a,b) Real (Re) and imaginary (Im) parts of the spin projection, ${\rm Re}[S_{y}^{\pm}]$ and ${\rm Im}[S_{y}^{\pm}]$, as a function of $k_{x}$ for distinct values of $k_{y}$ at $B=0$ where solid (dashed) curves correspond to the spin projections obtained with $\psi_{+}$ ($\psi_{-}$). Also,  light gray (brown) curves showing a sharp (smooth) transition across $k_{x}=0$ correspond to $\Gamma_{\uparrow,\downarrow}=0$ and $k_{y}=0$ ($k_{y}=0.5$). (c,d) Same as (a,b) but now as a function of $k_{y}$ at distinct values of $k_{x}$. (e,f) Real and imaginary parts of the spin-projection differences $\Delta S_{y}=S_{y}^{-}-S_{y}^{+}$ as a function of $B$ and $\gamma$ at  finite momenta $k_{x,y}=0.5$. The dashed cyan lines indicate the regimes where exceptional points occur, which then mark the ends of  ${\rm Re\Delta S_{y}}=0$ (uniform blue region). Parameters:  $\Gamma_{\uparrow}=3$, $\alpha=1$, $\Gamma_{\downarrow}=0$, $\mu=1$.	}
	\label{fig4}
\end{figure}

Having established the emergence of  exceptional degeneracies in the bulk of 2D Rashba semiconductors, now we turn our attention to how the spins here behave under non-Hermiticity. This is motivated by the fact that it is the spin  an important quantity for several phenomena in semiconductors, useful for spintronics and topological phenomena. In particular, in this part we focus on the spin expectation values, which here will be referred to as spin projections and are obtained by 
\begin{equation}
\label{spinProjections}
S^{\eta}_{j}=\Psi^{\dagger}_{\eta}\sigma_{j}\Psi_{\eta}
\end{equation}
where $\Psi^{\dagger}_{\eta}$ is given by Eq.\,(\ref{EvectorsNH}) with $\eta=\pm$ and $\sigma_{j}$ the $j$-th spin Pauli matrix. Thus, $S^{\eta}_{j}$ represents the spin projection along $j$ axis associated to $\eta=\pm$. By plugging Eq.\,(\ref{EvectorsNH})  into Eq.\,(\ref{spinProjections}), we obtain
\begin{equation}
\label{SxSy}
\begin{split}
S^{\pm}_{x}&=\frac{1}{2\alpha |k|^{2}}\left[k_{x}\gamma \pm k_{y} \sqrt{\alpha^{2}|k|^{2}-\gamma^{2}} \right]\,,\\
S^{\pm}_{y}&=\frac{1}{2\alpha |k|^{2}}\left[k_{y}\gamma \mp k_{x} \sqrt{\alpha^{2}|k|^{2}-\gamma^{2}} \right]\,,\\
\end{split}
\end{equation}
for the spin projections along $x$ and $y$, respectively, while $S^{\pm}_{z}=0$ along $z$. Note that here $|k|^{2}=k_{x}^{2}+k_{y}^{2}$ and $\gamma=(\Gamma_{\uparrow}-\Gamma_{\downarrow})/2$ characterizes the amount of non-Hermiticity due to the ferromagnet lead, see Eqs.\,(\ref{openNH}) and (\ref{NHSelf}). As before, the effect of the Zeeman field  along $x$ considered in Fig.\,\ref{fig1} can be included by replacing $\alpha k_{y}\rightarrow B+\alpha k_{y}$. The expressions given by  Eqs.\,(\ref{SxSy}) are relatively simple and permit us to  identify  the impact of non-Hermiticity on the spin projections by naked eye. In the Hermitian regime, when $\gamma=0$, the spin projections reduce to  $S^{\pm}_{x}=\pm k_{y}/|k|$ and $S^{\pm}_{y}=\mp k_{x}/|k|$, as expected \cite{winkler2003spin,chen2021spin}.  Note that   $S^{+}_{y(x)}$ and $S^{-}_{y(x)}$   change their sign when $k_{x(y)}$ varies from negative to positive values passing through $k_{x(y)}=0$, see  gray and pink curves in Fig.\,\ref{fig4}(a) showing the behaviour of $S^{+}_{y}$. The sign of $S^{\pm}_{y(x)}$  remains, however, upon variations of $k_{y(x)}$, as depicted in gray and pink curves in Fig.\,\ref{fig4}(c).

For finite non-Hermiticity, characterized by $\gamma\neq0$, the behaviour of the spin projections $S^{\pm}_{x (y)}$ is highly unusual. A finite $\gamma$ generates a linear in momentum term proportional to $k_{x(y)}\gamma$ for $S_{x(y)}$  and renormalizes the  Hermitian component with $\sqrt{\alpha^{2}|k|^{2}-\gamma^{2}}$, see first and second terms in Eqs.\,(\ref{SxSy}). Both terms reveal a unique effect of non-Hermiticity. The first part of $S^{\pm}_{x (y)}$, proportional to $k_{x(y)}\gamma$, is always real and  appears along the same direction of the spin projection, in contrast to the Hermitian contribution where $S^{\pm}_{x(y)}$ is only finite along $y(x)$. 
The second part of $S^{\pm}_{x (y)}$ is real for $|\alpha||k|>|\gamma|$, which then adds up to the first part, but becomes imaginary for $|\alpha||k|<|\gamma|$.  Thus, the appearance of an imaginary part in the spin projections for $|\alpha||k|<|\gamma|$ can be interpreted as a signal of their lifetime, which   becomes highly anisotropic in momentum space. At $\alpha^{2}|k|^{2}-\gamma^{2}=0$,  the second term in Eqs.\,(\ref{SxSy}) vanish and the spin projections $S^{\pm}_{j}$ coalesce, namely, they merge into a single value that is given by
 \begin{equation}
 \label{SxSyEP}
\begin{split}
S^{\pm,{\rm EP}}_{x(y)}=\frac{k_{x(y)}}{2|k|^{2}}\,.
 \end{split}
\end{equation}
Interestingly, the condition $\alpha^{2}|k|^{2}-\gamma^{2}=0$, which leads to this spin projection coalescence, is the same condition that determines the formation of exceptional degeneracies discussed in previous section, see Eq.\,(\ref{EPcondition}). Thus, the coalescence effect of $S^{\pm}_{j}$ can be seen as unique NH effect without analog in Hermitian systems.  We also note that  along the lines connecting  these exceptional degeneracies, which correspond to energy lines that resemble bulk Fermi arcs, the spin projections acquire a finite imaginary part with a natural physical interpretation as discussed in previous paragraph.

 %%%%%%%%%%%%%%%%%
%Fig 4
%%%%%%%%%%%%%%%%%%

In order to gain visual understanding of the spin projection coalescence, in Fig.\,\ref{fig4} we plot the real and imaginary parts of $S^{\pm}_{y}$ as a function of  momenta (a-d) and in the $B-\gamma$ plane (e,f). At $k_{y}=0$, the real part of the spin projections $S^{\pm}_{y}$ vanishes along a line of  $k_{x}$ and the ends of such line  mark the  EP momenta obtained from Eq.\,(\ref{EPcondition}) and 
given by $|k_{x}^{\rm EP}|=|\gamma|/|\alpha|$, see solid and dashed blue curves in Fig.\,\ref{fig4}(a); see also Eqs.\,(\ref{SxSy}) and (\ref{SxSyEP}). The imaginary part of $S^{\pm}_{y}$ undergoes a coalescence effect as well at the EP momenta $\pm k_{x}^{\rm EP}$ but  acquires large values between them and vanish at $k_{x}=0$ [Fig.\,\ref{fig4}(b)]. For $k_{y}>0$, the coalescence effect persists, with smaller imaginary parts, but the real part does not vanish anymore and, instead, develops a maximum  at $k_{x}=0$ favouring a large positive spin projection along $y$, see Eqs.\,(\ref{SxSy}) and (\ref{SxSyEP}). For $k_{y}<0$, the spin projection $S^{\pm}_{y}$ has instead a minimum at $k_{x}=0$, favouring a large negative spin projection along $y$. The coalescence of spin projections is also observed  in Fig.\,\ref{fig4}(c,d), where we plot the real and imaginary parts of $S^{\pm}_{y}$ as a function of $k_{y}$ at fixed values of $k_{x}$. At $k_{x}=0$, no EP transition is observed in $S^{\pm}_{y}$ because the square root term that  gives rise to EPs is multiplied by zero and hence vanishes, see  blue curves in Fig.\,\ref{fig4}(c,d)  and also Eqs.\,(\ref{SxSy}). However, for finite $k_{x}$, the spin projections develop a clear EP transition, revealing that their coalescence is a highly tunable NH bulk  effect.

The spin projection coalescence discussed above requires finite momenta,  Rashba SOC, and non-Hermiticity, a combination of ingredients inherent to NH Rashba semiconductors. Furthermore, it is also possible to tune and control the spin projections by an in-plane Zeeman field $B$, e.g., along $x$ as  sketched in Fig.\,\ref{fig1}, see also discussions below  Eqs.\,(\ref{Hclosed}) and (\ref{SxSy}). To support this idea, in Fig.\,\ref{fig4}(e,f) we present the Re and the Im parts of the difference between spin projection along $y$, ${\rm Re}\Delta S_{y}={\rm Re}(S^{-}_{y}-S^{+}_{y)}$ and ${\rm Im}\Delta S_{y}={\rm Im}(S^{-}_{y}-S^{+}_{y)}$,
 at fixed $k_{x,y}$ as a function of $B$ and $\gamma$. Here, the blue regions indicate ${\rm Re}\Delta S_{y}=0$ and ${\rm Im}\Delta S_{y}=0$ and their borders   show the exceptional degeneracies, indicated in cyan dashed curves. At fixed non-Hermiticity ($\gamma$), the Zeeman field $B$ induces the coalescence of spin projections $S_{y}^{\pm}$ at  EPs, which requires small (large) $B$ for weak (strong) non-Hermiticity. Therefore,  Zeeman fields offer another possibility for tuning and controlling the spin-projection coalescence at exceptional degeneracies in 2D Rashba semiconductors.

%%%%%%%%%%%%%%%%%%%%%%%%%%%%%%%
%SECTION IV:       SPECTRAL SIGNATURES                      %
%%%%%%%%%%%%%%%%%%%%%%%%%%%%%%%

\section{Spectral signatures}
\label{sec:sectionVII}
In this part we explore the spectral function as a potential way for detecting   EPs in NH Rashba semiconductors, which can be measured  by tools such as ARPES \cite{hufner2013photoelectron,lv2019angle,yu2020relevance,doi:10.7566/JPSJ.84.072001,RevModPhys.93.025006,PhysRevResearch.4.L022018}. The spectral function can be obtained as $A(k,\omega)=-{\rm Im Tr} {(G^{r}-G^{a})}$, where $G^{r}=(\omega-H_{\rm eff})^{-1}$  and $G^{a}=(G^{r})^{\dagger}$ are the retarded and advanced Green's functions, respectively \cite{mahan2013many,zagoskin}. The expressions for $G^{r(a)}$ are not complicated, which allows us to write down the spectral function as
\begin{equation}
A(\omega,k)=-2{\rm Im}\bigg[\frac{1}{\omega-E_{-}(k)}+\frac{1}{\omega-E_{+}(k)} \bigg]\,,
\end{equation}
where $E_{\pm}(k)$ are given by Eq.\,(\ref{EnergiesNH}). Although this expression already reveals the behaviour of the spectral function in our system, it is useful to write it as 
\begin{equation}
\label{A22}
 A(\omega,k)=-2{\rm Im}\sum_{i=\pm}\bigg[\frac{\omega-{\rm Re}E_{i}}{D(\omega,E_{i})} -i\frac{{\rm Im}E_{i}}{D(\omega,E_{i})} \bigg]\,,
 \end{equation}
with $D(\omega,E_{i})=(\omega-{\rm Re}E_{i})^{2}+({\rm Im}E_{i})^{2}$. Now, we clearly see that the spectral function in our system is a sum of two Lorentzians centered at $\omega={\rm Re}E_{\pm}$ with their height and width characterized by ${\rm Im}E_{\pm}$. It is thus evident in Eq.\,(\ref{A22}) that, at the EPs and at momenta between them where eigenvalues merge, instead of two Lorentzian resonances  we only have one.

\begin{figure}[!t]
	\centering
	\includegraphics[width=0.49\textwidth]{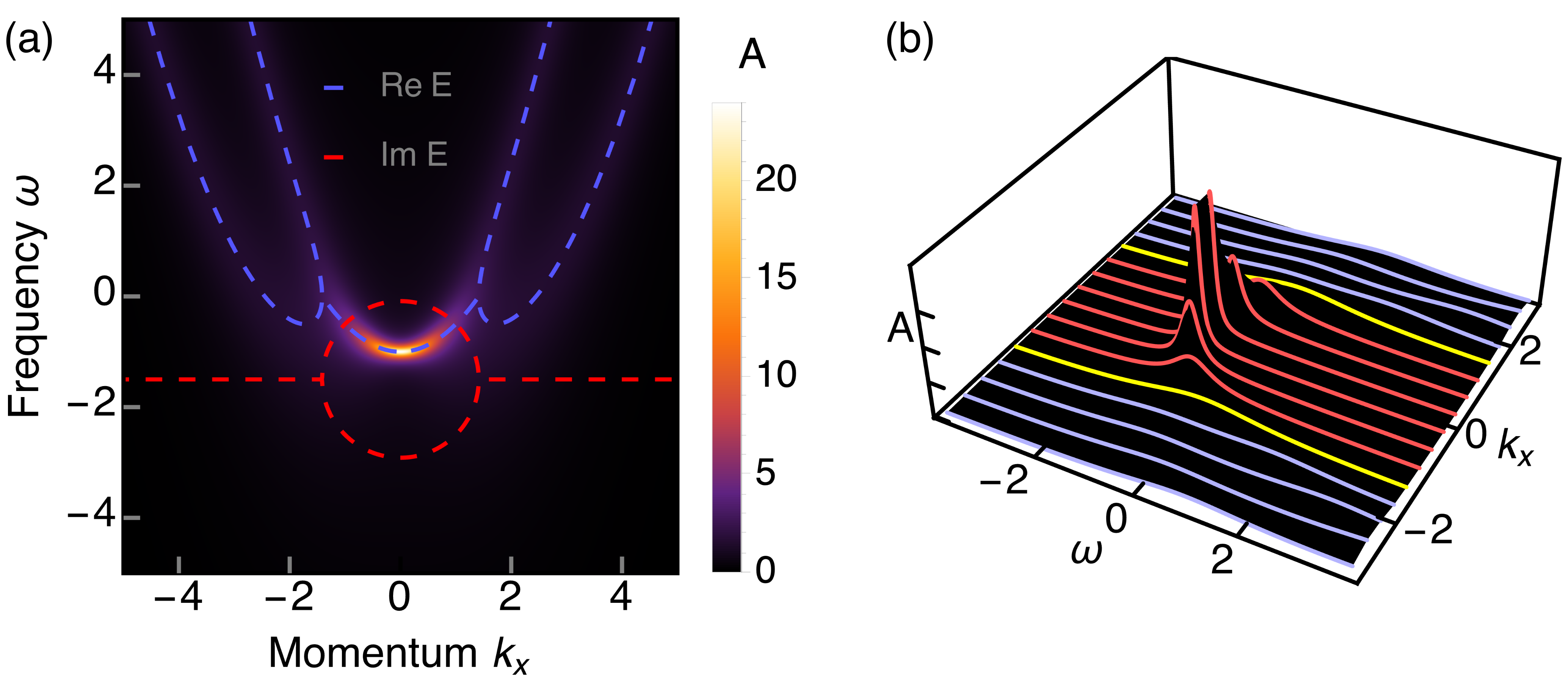} 
	\caption{(a) Spectral function $A(\omega,k)$ as a function of $\omega$ and momentum $k_{x}$. Here, also the real and imaginary parts of the eigenvalues are  shown in dashed blue  and dashed   red curves, respectively. (b) Line cuts of A with EPs (yellow), region between EPs (red), and region beyond (light blue).
	 Parameters:  $\Gamma_{\uparrow}=3$, $k_{y}=0$, $\alpha=1$, $\Gamma_{\downarrow}=0$, $\mu=1$, $B=0.5$.}
	\label{fig5}
\end{figure}

The behaviour of the spectral function can be further visualized in Fig.\,\ref{fig5}(a), where we plot $A$ as a function of $\omega$ and $k_{x}$ at $k_{y}=0$, $B=0.5$; the real and imaginary parts of the eigenvalues are  shown in dashed blue  and dashed   red curves.  We also plot line cuts in Fig.\,\ref{fig5}(b) for distinct $k$-values spanning the momenta at which EP occur (yellow curves), and momenta between EPs (red curves).  The immediate observation is that $A$ develops high intensity regions  for a line of momenta bounded by the EP momenta, revealing both the position of EPs and   the formation of the finite real energy Fermi arcs.  At the EPs, marked by yellow lines in Fig.\,\ref{fig5}(b), the spectral function undergoes a transition along $k$ from having two   resonances to having a single resonance centered at ${\rm Re}E_{1,2}\equiv\xi_{k}$, see also Eq.\,(\ref{A22}).  For momenta between the EPs, the real parts stick together ${\rm Re}E_{1,2}\equiv\xi_{k}$ and a single resonance remains all over these momenta, see   red curves. Interestingly, the resonances   between the EPs acquire very large values which could be also useful for  identifying the bulk Fermi arc at real energies found here. Before ending this part, we would like to mention that the length of the high intensity region seen in Fig.\,\ref{fig5}(a) along $k$, which is the length of the arc,  not only permits to estimate the EP momenta but it also allows to identify the amount of non-Hermiticity $\gamma$ according to Eq.\,(\ref{EPcondition}). Alternatively, this feature could provide a way to assess the strength of SOC $\alpha$, provided $\gamma$ is known as dictated by Eq.\,(\ref{EPcondition}). In sum, the spectral function reveals unique features of EPs and offers a powerful route for their detection.

 %%%%%%%%%%%%%%%%%
%Fig 5
%%%%%%%%%%%%%%%%%%

%%%%%%%%%%%%%%%%%%%%%%%%%%%%%%%%%%%%
% SECTION V:                        CONCLUSIONS                     %
%%%%%%%%%%%%%%%%%%%%%%%%%%%%%%%%%%%%

\section{Conclusions}
\label{sec:Concl}
We have demonstrated that the interplay between non-Hermiticity and Rashba spin-orbit coupling in semiconductors gives rise to the emergence of stable and highly tunable bulk exceptional degeneracies. We have found that these degeneracies form rings in two-dimensional momentum space and signal  the ends of lines  forming due to the coalescence of eigenvalues at finite real energy.  Interestingly, the lines at finite real energies have the appearance of  bulk Fermi arcs but now at finite energies, suggesting new possibilities for non-Hermitian bulk phenomena \cite{PhysRevResearch.4.L022064},  whose detail properties, however, deserve a proper investigation and will be addressed in a future study. We have also shown that the exceptional degeneracies and bulk Fermi arcs can be controlled by an in-plane Zeeman field, albeit larger non-Hermiticity values are then needed. Furthermore, we have discovered that the spin projections coalesce at the exceptional degeneracies and can easily achieve higher values than in the Hermitian regime.  We also demonstrate  that the exceptional degeneracies induce large spectral features along momenta connecting them, unique features that can be directly detected by ARPES. Taken together, the results presented here put semiconductors with Rashba SOC as an interesting arena for the realization of highly tunable non-Hermitian bulk  phenomena.

We also note that there is reasonable evidence suggesting that the system studied here as well as the main findings are within experimental reach. In fact, very similar systems as studied here have already been fabricated, including semiconductors with SOC made of InAs \cite{Kjaergaard,PhysRevB.93.155402,Suominen17,bottcher2018superconducting,fornieri2019evidence,o2021epitaxial} or InSb \cite{gazibegovic2019bottom,ke2019ballistic,xue2019gate,lei2019quantum,chen2021strong}  and ferromagnets producing sizeable Zeeman fields  \cite{katmis2016high,liu2019semiconductor,yang2020spin,vaitiekenas2021zero,escribano2022semiconductor,PhysRevB.105.L041304,razmadze2022supercurrent}. The coupling between ferromagnet and semiconductor, here characterized by $\Gamma_{\sigma}$, can be controlled by an appropriate manipulation of both the spin-dependent density of states in the lead and the tunneling between lead and semiconductor. While the Zeeman field of the ferromagnet enables a distinct density of states for different spins,  thus producing different couplings $\Gamma_{\sigma}$,  the overall strength of such couplings can be tuned by   inserting a normal     potential barrier of finite thickness between the semiconductor and ferromagnet lead, e.g., by using a few nm thick InGaAs layer \cite{PhysRevB.93.155402}.   We have estimated that, using $\mu_{\rm L}=1$\,meV, $t_{z}=2$\,meV, a ferromagnet with a Zeeman field    of the order of $3$meV  would be necessary to achieve the conditions of distinct couplings of $\Gamma_{\uparrow}=0.4$\,meV and $\Gamma_{\downarrow}=0$, thus giving $\gamma=0.2$\,meV.
Under these conditions and considering   $k_{y}=0$ and $\alpha=20$\,meVnm, which is in the range of  the SOC in InAs and InSb,   we obtain that the EP conditions are satisfied for $k_{x}\approx0.7$\,nm$^{-1}$ or $k_{x}^{-1}=150$\,nm, which is clearly in the range of reasonable length scales in these systems, such as the length scale related to SOC \cite{Lutchyn2018Majorana}.   We can thus conclude that, despite the possible challenges, there already exist experiments suggesting that the NH  semiconductor and the exceptional points studied here represent a feasible idea.

%%%%%%%%%%%%%%%%%%%%%%%%%%%%%%%
%                        ACKNOWLEDGMENTS                               %
%%%%%%%%%%%%%%%%%%%%%%%%%%%%%%%
\begin{acknowledgements}
We thank A. M. Black-Schaffer and P. Oppeneer for useful discussions. We acknowledge financial support from the Swedish Research Council  (Vetenskapsr\aa det Grant No.~2021-04121), the G\"{o}ran Gustafsson Foundation (Grant No.~2216), the Scandinavia-Japan Sasakawa Foundation  (Grant No.~GA22-SWE-0028), the Royal Swedish Academy of Sciences (Grant No.~PH2022-0003), and the  C.F. Liljewalchs stipendiestiftelse. 
 \end{acknowledgements}

%%%%%%%%%%%%%%%%%%%%%%%%%%%%%%%
%                                  APPENDIX                                  %
%%%%%%%%%%%%%%%%%%%%%%%%%%%%%%%
 
\appendix
\renewcommand{\thepage}{A\arabic{page}}
\setcounter{page}{1}
 \renewcommand{\thefigure}{A\arabic{figure}}
\setcounter{figure}{0}

\section{Derivation of the self-energy}
\label{Appendix}
In this part we show how the retarded self-energy $\Sigma^{r}$ given by Eq.\,(\ref{NHSelf}) is derived.
At this point, we   remind that our open system is modeled by an effective Hamiltonian $H_{\rm eff}$ given by Eq.\,(\ref{openNH})  that contains the Hamiltonian $H_{\rm R}$ of an isolated 2D semiconductor and a self-energy due to its coupling to a semi-infinite ferromagnetic lead, modeled by the Hamiltonian $H_{\rm L}$. Thus, the open system we study   consists of a  2D junction along z: the closed system $H_{\rm R}$ corresponds to a 2D semiconductor, which can be seen of as having one site along $z$, while the ferromagnetic lead is semi-infinite along $z$ for negative $z$, see Fig.\,\ref{fig1}.

As already pointed out in Section \ref{sec:Model}, the self-energy $\Sigma^{r}$ depends on frequency under general circumstances and can be obtained as $\Sigma^{r}(\omega)=V^{\dagger}g^{r}_{\rm L}(\omega)V$, where $g_{\rm L}^{r}(\omega)=(\omega-H_{\rm L})^{-1}$   is the retarded Green’s function of the semi-infinite lead and $V$ is the hopping matrix between the system and the lead. Because $V$ is only finite between the nearest neighbor sites of the ferromagnet lead and the Rashba semiconductor, it is possible to project the self-energy onto $H_{\rm R}$, which reads
\begin{equation}
\label{ZZZ}
\Sigma^{r}_{1_{\rm R}1_{\rm R}}(\omega)=\bra{1_{\rm R}}V^{\dagger}\ket{1_{\rm L}}\bra{1_{\rm L}}g^{r}_{\rm L}(\omega)\ket{1_{\rm L}}\bra{1_{\rm L}}V\ket{1_{\rm R}}\,,
\end{equation}
where $1_{\rm L}$ denotes the first site of the lead, closest to the semiconductor, while $1_{\rm R}$ denotes the only site in the Rashba semiconductor along   the $z$-direction. We also set $\bra{1_{\rm L}}V\ket{1_{\rm R}}\equiv V_{1_{\rm L}1_{\rm R}}=-t'\sigma_{0}$, where $t'$ is the hopping amplitude between sites $1_{\rm L}$ in the lead and site $1_{\rm R}$ in the semiconductor. It is interesting to notice that Eq.\,(\ref{ZZZ}) implies that it is only required  the Green's function of the lead at site $1_{\rm L}$, which is  the surface lead Green's function.  

To find the surface lead Green's function, it is important to remark that as a result of the lead being semi-infinite along the negative $z$-direction, it contains an infinite number of sites in this direction, with the   Hamiltonian for each site ($i_{\rm L}$) given by the on-site terms as $[H_{\rm L}]_{i_{\rm L}i_{\rm L}}=\xi_{k}^{\rm L}\sigma_{0} +B_{\rm L}\sigma_{z} $, where $\xi_{k}^{\rm L}=\hbar^{2}k^{2}/2m-\mu_{\rm L}$ is the kinetic term in the lead,  with $k=(k_{x},k_{y})$, and chemical potential $\mu_{\rm L}$.  Also, $B_{\rm L}$ is the Zeeman energy, which appears because the lead is ferromagnetic, but it might be also due to    an external magnetic field. Thus, to find the lead Green's function, it is important to take into account that 
  $H_{\rm L}$ is an infinite matrix. Then, by using a recursive approach \cite{PhysRevB.91.024514}, we   find   $ \bra{1_{\rm L}}g^{r}_{\rm L}(\omega)\ket{1_{\rm L}} ={\rm diag}(g^{r}_{\uparrow\uparrow},g^{r}_{\downarrow\downarrow})$, with the diagonal entries given by
 \begin{equation}
 g^{r}_{\sigma\sigma}(\omega)=\frac{1}{|t_{z}|}\left[\frac{\omega-\epsilon_{\sigma}}{2|t_{z}|}- i\sqrt{1\,-\,\left(\frac{\omega-\epsilon_{\sigma}}{2|t_{z}|}\right)^{2}}\right]\,.
 \end{equation}
 for  $|(\omega-\epsilon^{e(h)}_{\sigma})/2|t_{z}||<1$. Here, $\epsilon_{\sigma}=\xi^{\rm L}_{k}+\sigma B_{\rm L}$. We can then write the self-energy in spin space as
\begin{equation}
\label{SSSS}
\Sigma_{1_{\rm R}1_{\rm R}}^{r}=\begin{pmatrix}
t'^{2}g_{\uparrow\uparrow}^{r}&0\\
0&t'^{2}g_{\downarrow\downarrow}^{r}
\end{pmatrix},
\end{equation}
which    involves real and imaginary terms due to the imaginary part of the lead Green's function. While under general circumstances, both real and imaginary parts impact the Hamiltonian of the 2D Rashba semiconductor $H_{\rm R}$, only the imaginary term gives rise to NH physics. 

Here, we are here interested in the effect of the NH part of the self-energy and for this purpose we  
  it is useful to carry out some approximations. First, we consider   $|(\omega-\epsilon^{e(h)}_{\sigma})/(2t_{z})|\ll |(\mu_{\rm L}-\sigma B)/(2t_{z})|<1$, which can be seen to be a sort of wide band limit widely used in quantum transport \cite{datta1997electronic,Ryndyk2009} that permits us to neglect the dependence on both frequency and momentum  in the lead Green's function $g_{\rm L}^{r}$. Notice that we still assume values of $B$ and $\mu_{\rm L}$ to be large enough such that $g_{\rm L}^{r}$ develops   different imaginary terms  for different spins. Second, we neglect the real part of the self-energy as it only introduces shifts into the semiconductor Hamiltonian $H_{\rm R}$: we only keep the imaginary term of the self-energy because it renders NH the  effective Hamiltonian $H_{\rm eff}$ of the total system. The possible energy shifts due to the real part of the self-energy can be incorporated by an appropriate renormalization of the semiconductor Hamiltonian  by spanning over relevant parameter regimes. We can thus safely neglect the frequency and momentum dependence of the self-energy and also its real part, enabling us to only focus on the imaginary component for inducing NH physics and EPs. Hence, we can approximate the self-energy $\Sigma_{1_{\rm R}1_{\rm R}}^{r}$ in Eq.\,(\ref{SSSS}) as
\begin{equation}
\label{selfenergyEq2}
\Sigma^{r}(\omega=0)=-i \Gamma \sigma_{0}-i\gamma \sigma_{z}\,,
\end{equation}
where $\Gamma=(\Gamma_{\uparrow}+\Gamma_{\downarrow})/2$ and $\gamma=(\Gamma_{\uparrow}-\Gamma_{\downarrow})/2$. Here we have defined $\Gamma_{\sigma}=\pi|t'|^{2}\rho_{\rm L}^{\sigma}$, where $\rho_{\rm L}^{\uparrow(\downarrow)}=[1/(t_{z}\pi)]\sqrt{1-[(\mu_{\rm L}\mp B)/(2t_{z})]^{2}}$  is the spin-polarized surface density of states of the lead. Eq.\,(\ref{selfenergyEq2}) is presented as Eq.\,(\ref{NHSelf}) of the main text, where its impact for inducing EPs is further discussed.
 
%%%%%%%%%%%%%%%%%%%%%%%%%%%%%%
%         REFERENCES        %
%%%%%%%%%%%%%%%%%%%%%%%%%%%%%%%
\bibliography{biblio}

\begin{thebibliography}{90}%
\makeatletter
\providecommand \@ifxundefined [1]{%
 \@ifx{#1\undefined}
}%
\providecommand \@ifnum [1]{%
 \ifnum #1\expandafter \@firstoftwo
 \else \expandafter \@secondoftwo
 \fi
}%
\providecommand \@ifx [1]{%
 \ifx #1\expandafter \@firstoftwo
 \else \expandafter \@secondoftwo
 \fi
}%
\providecommand \natexlab [1]{#1}%
\providecommand \enquote  [1]{``#1''}%
\providecommand \bibnamefont  [1]{#1}%
\providecommand \bibfnamefont [1]{#1}%
\providecommand \citenamefont [1]{#1}%
\providecommand \href@noop [0]{\@secondoftwo}%
\providecommand \href [0]{\begingroup \@sanitize@url \@href}%
\providecommand \@href[1]{\@@startlink{#1}\@@href}%
\providecommand \@@href[1]{\endgroup#1\@@endlink}%
\providecommand \@sanitize@url [0]{\catcode `\\12\catcode `\$12\catcode
  `\&12\catcode `\#12\catcode `\^12\catcode `\_12\catcode `\%12\relax}%
\providecommand \@@startlink[1]{}%
\providecommand \@@endlink[0]{}%
\providecommand \url  [0]{\begingroup\@sanitize@url \@url }%
\providecommand \@url [1]{\endgroup\@href {#1}{\urlprefix }}%
\providecommand \urlprefix  [0]{URL }%
\providecommand \Eprint [0]{\href }%
\providecommand \doibase [0]{https://doi.org/}%
\providecommand \selectlanguage [0]{\@gobble}%
\providecommand \bibinfo  [0]{\@secondoftwo}%
\providecommand \bibfield  [0]{\@secondoftwo}%
\providecommand \translation [1]{[#1]}%
\providecommand \BibitemOpen [0]{}%
\providecommand \bibitemStop [0]{}%
\providecommand \bibitemNoStop [0]{.\EOS\space}%
\providecommand \EOS [0]{\spacefactor3000\relax}%
\providecommand \BibitemShut  [1]{\csname bibitem#1\endcsname}%
\let\auto@bib@innerbib\@empty
%</preamble>
\bibitem [{\citenamefont {El-Ganainy}\ \emph {et~al.}(2018)\citenamefont
  {El-Ganainy}, \citenamefont {Makris}, \citenamefont {Khajavikhan},
  \citenamefont {Musslimani}, \citenamefont {Rotter},\ and\ \citenamefont
  {Christodoulides}}]{el2018non}%
  \BibitemOpen
  \bibfield  {author} {\bibinfo {author} {\bibfnamefont {R.}~\bibnamefont
  {El-Ganainy}}, \bibinfo {author} {\bibfnamefont {K.~G.}\ \bibnamefont
  {Makris}}, \bibinfo {author} {\bibfnamefont {M.}~\bibnamefont {Khajavikhan}},
  \bibinfo {author} {\bibfnamefont {Z.~H.}\ \bibnamefont {Musslimani}},
  \bibinfo {author} {\bibfnamefont {S.}~\bibnamefont {Rotter}},\ and\ \bibinfo
  {author} {\bibfnamefont {D.~N.}\ \bibnamefont {Christodoulides}},\ }\bibfield
   {title} {\bibinfo {title} {Non-hermitian physics and pt symmetry},\ }\href
  {https://www.nature.com/articles/nphys4323} {\bibfield  {journal} {\bibinfo
  {journal} {Nat. Phys.}\ }\textbf {\bibinfo {volume} {14}},\ \bibinfo {pages}
  {11} (\bibinfo {year} {2018})}\BibitemShut {NoStop}%
\bibitem [{\citenamefont {{\"O}zdemir}\ \emph {et~al.}(2019)\citenamefont
  {{\"O}zdemir}, \citenamefont {Rotter}, \citenamefont {Nori},\ and\
  \citenamefont {Yang}}]{ozdemir2019parity}%
  \BibitemOpen
  \bibfield  {author} {\bibinfo {author} {\bibfnamefont {{\c{S}}.~K.}\
  \bibnamefont {{\"O}zdemir}}, \bibinfo {author} {\bibfnamefont
  {S.}~\bibnamefont {Rotter}}, \bibinfo {author} {\bibfnamefont
  {F.}~\bibnamefont {Nori}},\ and\ \bibinfo {author} {\bibfnamefont
  {L.}~\bibnamefont {Yang}},\ }\bibfield  {title} {\bibinfo {title}
  {Parity--time symmetry and exceptional points in photonics},\ }\href
  {https://www.nature.com/articles/s41563-019-0304-9} {\bibfield  {journal}
  {\bibinfo  {journal} {Nat. Mater.}\ }\textbf {\bibinfo {volume} {18}},\
  \bibinfo {pages} {783} (\bibinfo {year} {2019})}\BibitemShut {NoStop}%
\bibitem [{\citenamefont {Bergholtz}\ \emph {et~al.}(2021)\citenamefont
  {Bergholtz}, \citenamefont {Budich},\ and\ \citenamefont
  {Kunst}}]{RevModPhys.93.015005}%
  \BibitemOpen
  \bibfield  {author} {\bibinfo {author} {\bibfnamefont {E.~J.}\ \bibnamefont
  {Bergholtz}}, \bibinfo {author} {\bibfnamefont {J.~C.}\ \bibnamefont
  {Budich}},\ and\ \bibinfo {author} {\bibfnamefont {F.~K.}\ \bibnamefont
  {Kunst}},\ }\bibfield  {title} {\bibinfo {title} {Exceptional topology of
  non-hermitian systems},\ }\href
  {https://doi.org/10.1103/RevModPhys.93.015005} {\bibfield  {journal}
  {\bibinfo  {journal} {Rev. Mod. Phys.}\ }\textbf {\bibinfo {volume} {93}},\
  \bibinfo {pages} {015005} (\bibinfo {year} {2021})}\BibitemShut {NoStop}%
\bibitem [{\citenamefont {Ashida}\ \emph {et~al.}(2020)\citenamefont {Ashida},
  \citenamefont {Gong},\ and\ \citenamefont
  {Ueda}}]{doi:10.1080/00018732.2021.1876991}%
  \BibitemOpen
  \bibfield  {author} {\bibinfo {author} {\bibfnamefont {Y.}~\bibnamefont
  {Ashida}}, \bibinfo {author} {\bibfnamefont {Z.}~\bibnamefont {Gong}},\ and\
  \bibinfo {author} {\bibfnamefont {M.}~\bibnamefont {Ueda}},\ }\bibfield
  {title} {\bibinfo {title} {Non-hermitian physics},\ }\href
  {https://doi.org/10.1080/00018732.2021.1876991} {\bibfield  {journal}
  {\bibinfo  {journal} {Adv. Phys.}\ }\textbf {\bibinfo {volume} {69}},\
  \bibinfo {pages} {249} (\bibinfo {year} {2020})}\BibitemShut {NoStop}%
\bibitem [{\citenamefont {Parto}\ \emph {et~al.}(2020)\citenamefont {Parto},
  \citenamefont {Liu}, \citenamefont {Bahari}, \citenamefont {Khajavikhan},\
  and\ \citenamefont {Christodoulides}}]{parto2020non}%
  \BibitemOpen
  \bibfield  {author} {\bibinfo {author} {\bibfnamefont {M.}~\bibnamefont
  {Parto}}, \bibinfo {author} {\bibfnamefont {Y.~G.}\ \bibnamefont {Liu}},
  \bibinfo {author} {\bibfnamefont {B.}~\bibnamefont {Bahari}}, \bibinfo
  {author} {\bibfnamefont {M.}~\bibnamefont {Khajavikhan}},\ and\ \bibinfo
  {author} {\bibfnamefont {D.~N.}\ \bibnamefont {Christodoulides}},\ }\bibfield
   {title} {\bibinfo {title} {Non-hermitian and topological photonics: optics
  at an exceptional point},\ }\href
  {https://www.degruyter.com/document/doi/10.1515/nanoph-2020-0434/html?lang=en}
  {\bibfield  {journal} {\bibinfo  {journal} {Nanophotonics}\ }\textbf
  {\bibinfo {volume} {10}},\ \bibinfo {pages} {403} (\bibinfo {year}
  {2020})}\BibitemShut {NoStop}%
\bibitem [{\citenamefont {Wiersig}(2020)}]{wiersig2020review}%
  \BibitemOpen
  \bibfield  {author} {\bibinfo {author} {\bibfnamefont {J.}~\bibnamefont
  {Wiersig}},\ }\bibfield  {title} {\bibinfo {title} {Review of exceptional
  point-based sensors},\ }\href
  {https://opg.optica.org/prj/fulltext.cfm?uri=prj-8-9-1457&id=434541}
  {\bibfield  {journal} {\bibinfo  {journal} {Photonics Res.}\ }\textbf
  {\bibinfo {volume} {8}},\ \bibinfo {pages} {1457} (\bibinfo {year}
  {2020})}\BibitemShut {NoStop}%
\bibitem [{\citenamefont {Ding}\ \emph {et~al.}(2022)\citenamefont {Ding},
  \citenamefont {Fang},\ and\ \citenamefont {Ma}}]{ding2022non}%
  \BibitemOpen
  \bibfield  {author} {\bibinfo {author} {\bibfnamefont {K.}~\bibnamefont
  {Ding}}, \bibinfo {author} {\bibfnamefont {C.}~\bibnamefont {Fang}},\ and\
  \bibinfo {author} {\bibfnamefont {G.}~\bibnamefont {Ma}},\ }\bibfield
  {title} {\bibinfo {title} {Non-hermitian topology and exceptional-point
  geometries},\ }\href {https://www.nature.com/articles/s42254-022-00516-5}
  {\bibfield  {journal} {\bibinfo  {journal} {Nat. Rev. Phys.}\ ,\ \bibinfo
  {pages} {745}} (\bibinfo {year} {2022})}\BibitemShut {NoStop}%
\bibitem [{\citenamefont {Gong}\ \emph {et~al.}(2018)\citenamefont {Gong},
  \citenamefont {Ashida}, \citenamefont {Kawabata}, \citenamefont {Takasan},
  \citenamefont {Higashikawa},\ and\ \citenamefont {Ueda}}]{PhysRevX.8.031079}%
  \BibitemOpen
  \bibfield  {author} {\bibinfo {author} {\bibfnamefont {Z.}~\bibnamefont
  {Gong}}, \bibinfo {author} {\bibfnamefont {Y.}~\bibnamefont {Ashida}},
  \bibinfo {author} {\bibfnamefont {K.}~\bibnamefont {Kawabata}}, \bibinfo
  {author} {\bibfnamefont {K.}~\bibnamefont {Takasan}}, \bibinfo {author}
  {\bibfnamefont {S.}~\bibnamefont {Higashikawa}},\ and\ \bibinfo {author}
  {\bibfnamefont {M.}~\bibnamefont {Ueda}},\ }\bibfield  {title} {\bibinfo
  {title} {Topological phases of non-hermitian systems},\ }\href
  {https://doi.org/10.1103/PhysRevX.8.031079} {\bibfield  {journal} {\bibinfo
  {journal} {Phys. Rev. X}\ }\textbf {\bibinfo {volume} {8}},\ \bibinfo {pages}
  {031079} (\bibinfo {year} {2018})}\BibitemShut {NoStop}%
\bibitem [{\citenamefont {Zhou}\ and\ \citenamefont
  {Lee}(2019)}]{PhysRevB.99.235112}%
  \BibitemOpen
  \bibfield  {author} {\bibinfo {author} {\bibfnamefont {H.}~\bibnamefont
  {Zhou}}\ and\ \bibinfo {author} {\bibfnamefont {J.~Y.}\ \bibnamefont {Lee}},\
  }\bibfield  {title} {\bibinfo {title} {Periodic table for topological bands
  with non-hermitian symmetries},\ }\href
  {https://doi.org/10.1103/PhysRevB.99.235112} {\bibfield  {journal} {\bibinfo
  {journal} {Phys. Rev. B}\ }\textbf {\bibinfo {volume} {99}},\ \bibinfo
  {pages} {235112} (\bibinfo {year} {2019})}\BibitemShut {NoStop}%
\bibitem [{\citenamefont {Kawabata}\ \emph
  {et~al.}(2019{\natexlab{a}})\citenamefont {Kawabata}, \citenamefont
  {Shiozaki}, \citenamefont {Ueda},\ and\ \citenamefont
  {Sato}}]{PhysRevX.9.041015}%
  \BibitemOpen
  \bibfield  {author} {\bibinfo {author} {\bibfnamefont {K.}~\bibnamefont
  {Kawabata}}, \bibinfo {author} {\bibfnamefont {K.}~\bibnamefont {Shiozaki}},
  \bibinfo {author} {\bibfnamefont {M.}~\bibnamefont {Ueda}},\ and\ \bibinfo
  {author} {\bibfnamefont {M.}~\bibnamefont {Sato}},\ }\bibfield  {title}
  {\bibinfo {title} {Symmetry and topology in non-hermitian physics},\ }\href
  {https://doi.org/10.1103/PhysRevX.9.041015} {\bibfield  {journal} {\bibinfo
  {journal} {Phys. Rev. X}\ }\textbf {\bibinfo {volume} {9}},\ \bibinfo {pages}
  {041015} (\bibinfo {year} {2019}{\natexlab{a}})}\BibitemShut {NoStop}%
\bibitem [{\citenamefont {Kato}(1966)}]{TKato}%
  \BibitemOpen
  \bibfield  {author} {\bibinfo {author} {\bibfnamefont {T.}~\bibnamefont
  {Kato}},\ }\href@noop {} {\emph {\bibinfo {title} {Perturbation theory of
  linear operators}}}\ (\bibinfo  {publisher} {Springer, New York},\ \bibinfo
  {year} {1966})\BibitemShut {NoStop}%
\bibitem [{\citenamefont {Heiss}(2004)}]{heiss2004exceptional}%
  \BibitemOpen
  \bibfield  {author} {\bibinfo {author} {\bibfnamefont {W.}~\bibnamefont
  {Heiss}},\ }\bibfield  {title} {\bibinfo {title} {Exceptional points--their
  universal occurrence and their physical significance},\ }\href
  {https://link.springer.com/article/10.1023/B:CJOP.0000044009.17264.dc}
  {\bibfield  {journal} {\bibinfo  {journal} {Czechoslov. J. Phys.}\ }\textbf
  {\bibinfo {volume} {54}},\ \bibinfo {pages} {1091} (\bibinfo {year}
  {2004})}\BibitemShut {NoStop}%
\bibitem [{\citenamefont {Berry}(2004)}]{berry2004physics}%
  \BibitemOpen
  \bibfield  {author} {\bibinfo {author} {\bibfnamefont {M.~V.}\ \bibnamefont
  {Berry}},\ }\bibfield  {title} {\bibinfo {title} {Physics of nonhermitian
  degeneracies},\ }\href
  {https://link.springer.com/article/10.1023/B:CJOP.0000044002.05657.04}
  {\bibfield  {journal} {\bibinfo  {journal} {Czechoslov. J. Phys.}\ }\textbf
  {\bibinfo {volume} {54}},\ \bibinfo {pages} {1039} (\bibinfo {year}
  {2004})}\BibitemShut {NoStop}%
\bibitem [{\citenamefont {Heiss}(2012)}]{Heiss_2012}%
  \BibitemOpen
  \bibfield  {author} {\bibinfo {author} {\bibfnamefont {W.~D.}\ \bibnamefont
  {Heiss}},\ }\bibfield  {title} {\bibinfo {title} {The physics of exceptional
  points},\ }\href {https://doi.org/10.1088/1751-8113/45/44/444016} {\bibfield
  {journal} {\bibinfo  {journal} {J. Phys. A Math. Theor.}\ }\textbf {\bibinfo
  {volume} {45}},\ \bibinfo {pages} {444016} (\bibinfo {year}
  {2012})}\BibitemShut {NoStop}%
\bibitem [{\citenamefont {Dembowski}\ \emph {et~al.}(2001)\citenamefont
  {Dembowski}, \citenamefont {Gr\"af}, \citenamefont {Harney}, \citenamefont
  {Heine}, \citenamefont {Heiss}, \citenamefont {Rehfeld},\ and\ \citenamefont
  {Richter}}]{PhysRevLett.86.787}%
  \BibitemOpen
  \bibfield  {author} {\bibinfo {author} {\bibfnamefont {C.}~\bibnamefont
  {Dembowski}}, \bibinfo {author} {\bibfnamefont {H.-D.}\ \bibnamefont
  {Gr\"af}}, \bibinfo {author} {\bibfnamefont {H.~L.}\ \bibnamefont {Harney}},
  \bibinfo {author} {\bibfnamefont {A.}~\bibnamefont {Heine}}, \bibinfo
  {author} {\bibfnamefont {W.~D.}\ \bibnamefont {Heiss}}, \bibinfo {author}
  {\bibfnamefont {H.}~\bibnamefont {Rehfeld}},\ and\ \bibinfo {author}
  {\bibfnamefont {A.}~\bibnamefont {Richter}},\ }\bibfield  {title} {\bibinfo
  {title} {Experimental observation of the topological structure of exceptional
  points},\ }\href {https://doi.org/10.1103/PhysRevLett.86.787} {\bibfield
  {journal} {\bibinfo  {journal} {Phys. Rev. Lett.}\ }\textbf {\bibinfo
  {volume} {86}},\ \bibinfo {pages} {787} (\bibinfo {year} {2001})}\BibitemShut
  {NoStop}%
\bibitem [{\citenamefont {Lee}\ \emph {et~al.}(2009)\citenamefont {Lee},
  \citenamefont {Yang}, \citenamefont {Moon}, \citenamefont {Lee},
  \citenamefont {Shim}, \citenamefont {Kim}, \citenamefont {Lee},\ and\
  \citenamefont {An}}]{PhysRevLett.103.134101}%
  \BibitemOpen
  \bibfield  {author} {\bibinfo {author} {\bibfnamefont {S.-B.}\ \bibnamefont
  {Lee}}, \bibinfo {author} {\bibfnamefont {J.}~\bibnamefont {Yang}}, \bibinfo
  {author} {\bibfnamefont {S.}~\bibnamefont {Moon}}, \bibinfo {author}
  {\bibfnamefont {S.-Y.}\ \bibnamefont {Lee}}, \bibinfo {author} {\bibfnamefont
  {J.-B.}\ \bibnamefont {Shim}}, \bibinfo {author} {\bibfnamefont {S.~W.}\
  \bibnamefont {Kim}}, \bibinfo {author} {\bibfnamefont {J.-H.}\ \bibnamefont
  {Lee}},\ and\ \bibinfo {author} {\bibfnamefont {K.}~\bibnamefont {An}},\
  }\bibfield  {title} {\bibinfo {title} {Observation of an exceptional point in
  a chaotic optical microcavity},\ }\href
  {https://doi.org/10.1103/PhysRevLett.103.134101} {\bibfield  {journal}
  {\bibinfo  {journal} {Phys. Rev. Lett.}\ }\textbf {\bibinfo {volume} {103}},\
  \bibinfo {pages} {134101} (\bibinfo {year} {2009})}\BibitemShut {NoStop}%
\bibitem [{\citenamefont {Choi}\ \emph {et~al.}(2010)\citenamefont {Choi},
  \citenamefont {Kang}, \citenamefont {Lim}, \citenamefont {Kim}, \citenamefont
  {Kim}, \citenamefont {Lee},\ and\ \citenamefont
  {An}}]{PhysRevLett.104.153601}%
  \BibitemOpen
  \bibfield  {author} {\bibinfo {author} {\bibfnamefont {Y.}~\bibnamefont
  {Choi}}, \bibinfo {author} {\bibfnamefont {S.}~\bibnamefont {Kang}}, \bibinfo
  {author} {\bibfnamefont {S.}~\bibnamefont {Lim}}, \bibinfo {author}
  {\bibfnamefont {W.}~\bibnamefont {Kim}}, \bibinfo {author} {\bibfnamefont
  {J.-R.}\ \bibnamefont {Kim}}, \bibinfo {author} {\bibfnamefont {J.-H.}\
  \bibnamefont {Lee}},\ and\ \bibinfo {author} {\bibfnamefont {K.}~\bibnamefont
  {An}},\ }\bibfield  {title} {\bibinfo {title} {Quasieigenstate coalescence in
  an atom-cavity quantum composite},\ }\href
  {https://doi.org/10.1103/PhysRevLett.104.153601} {\bibfield  {journal}
  {\bibinfo  {journal} {Phys. Rev. Lett.}\ }\textbf {\bibinfo {volume} {104}},\
  \bibinfo {pages} {153601} (\bibinfo {year} {2010})}\BibitemShut {NoStop}%
\bibitem [{\citenamefont {Gao}\ \emph {et~al.}(2015)\citenamefont {Gao},
  \citenamefont {Estrecho}, \citenamefont {Bliokh}, \citenamefont {Liew},
  \citenamefont {Fraser}, \citenamefont {Brodbeck}, \citenamefont {Kamp},
  \citenamefont {Schneider}, \citenamefont {H{\"o}fling}, \citenamefont
  {Yamamoto} \emph {et~al.}}]{gao2015observation}%
  \BibitemOpen
  \bibfield  {author} {\bibinfo {author} {\bibfnamefont {T.}~\bibnamefont
  {Gao}}, \bibinfo {author} {\bibfnamefont {E.}~\bibnamefont {Estrecho}},
  \bibinfo {author} {\bibfnamefont {K.}~\bibnamefont {Bliokh}}, \bibinfo
  {author} {\bibfnamefont {T.}~\bibnamefont {Liew}}, \bibinfo {author}
  {\bibfnamefont {M.}~\bibnamefont {Fraser}}, \bibinfo {author} {\bibfnamefont
  {S.}~\bibnamefont {Brodbeck}}, \bibinfo {author} {\bibfnamefont
  {M.}~\bibnamefont {Kamp}}, \bibinfo {author} {\bibfnamefont {C.}~\bibnamefont
  {Schneider}}, \bibinfo {author} {\bibfnamefont {S.}~\bibnamefont
  {H{\"o}fling}}, \bibinfo {author} {\bibfnamefont {Y.}~\bibnamefont
  {Yamamoto}}, \emph {et~al.},\ }\bibfield  {title} {\bibinfo {title}
  {Observation of non-hermitian degeneracies in a chaotic exciton-polariton
  billiard},\ }\href {https://www.nature.com/articles/nature15522} {\bibfield
  {journal} {\bibinfo  {journal} {Nature}\ }\textbf {\bibinfo {volume} {526}},\
  \bibinfo {pages} {554} (\bibinfo {year} {2015})}\BibitemShut {NoStop}%
\bibitem [{\citenamefont {Doppler}\ \emph {et~al.}(2016)\citenamefont
  {Doppler}, \citenamefont {Mailybaev}, \citenamefont {B{\"o}hm}, \citenamefont
  {Kuhl}, \citenamefont {Girschik}, \citenamefont {Libisch}, \citenamefont
  {Milburn}, \citenamefont {Rabl}, \citenamefont {Moiseyev},\ and\
  \citenamefont {Rotter}}]{doppler2016dynamically}%
  \BibitemOpen
  \bibfield  {author} {\bibinfo {author} {\bibfnamefont {J.}~\bibnamefont
  {Doppler}}, \bibinfo {author} {\bibfnamefont {A.~A.}\ \bibnamefont
  {Mailybaev}}, \bibinfo {author} {\bibfnamefont {J.}~\bibnamefont {B{\"o}hm}},
  \bibinfo {author} {\bibfnamefont {U.}~\bibnamefont {Kuhl}}, \bibinfo {author}
  {\bibfnamefont {A.}~\bibnamefont {Girschik}}, \bibinfo {author}
  {\bibfnamefont {F.}~\bibnamefont {Libisch}}, \bibinfo {author} {\bibfnamefont
  {T.~J.}\ \bibnamefont {Milburn}}, \bibinfo {author} {\bibfnamefont
  {P.}~\bibnamefont {Rabl}}, \bibinfo {author} {\bibfnamefont {N.}~\bibnamefont
  {Moiseyev}},\ and\ \bibinfo {author} {\bibfnamefont {S.}~\bibnamefont
  {Rotter}},\ }\bibfield  {title} {\bibinfo {title} {Dynamically encircling an
  exceptional point for asymmetric mode switching},\ }\href
  {https://www.nature.com/articles/nature18605} {\bibfield  {journal} {\bibinfo
   {journal} {Nature}\ }\textbf {\bibinfo {volume} {537}},\ \bibinfo {pages}
  {76} (\bibinfo {year} {2016})}\BibitemShut {NoStop}%
\bibitem [{\citenamefont {Yoshida}\ \emph {et~al.}(2019)\citenamefont
  {Yoshida}, \citenamefont {Peters}, \citenamefont {Kawakami},\ and\
  \citenamefont {Hatsugai}}]{PhysRevB.99.121101}%
  \BibitemOpen
  \bibfield  {author} {\bibinfo {author} {\bibfnamefont {T.}~\bibnamefont
  {Yoshida}}, \bibinfo {author} {\bibfnamefont {R.}~\bibnamefont {Peters}},
  \bibinfo {author} {\bibfnamefont {N.}~\bibnamefont {Kawakami}},\ and\
  \bibinfo {author} {\bibfnamefont {Y.}~\bibnamefont {Hatsugai}},\ }\bibfield
  {title} {\bibinfo {title} {Symmetry-protected exceptional rings in
  two-dimensional correlated systems with chiral symmetry},\ }\href
  {https://doi.org/10.1103/PhysRevB.99.121101} {\bibfield  {journal} {\bibinfo
  {journal} {Phys. Rev. B}\ }\textbf {\bibinfo {volume} {99}},\ \bibinfo
  {pages} {121101} (\bibinfo {year} {2019})}\BibitemShut {NoStop}%
\bibitem [{\citenamefont {Arouca}\ \emph {et~al.}(2022)\citenamefont {Arouca},
  \citenamefont {Cayao},\ and\ \citenamefont
  {Black-Schaffer}}]{arouca2022exceptionally}%
  \BibitemOpen
  \bibfield  {author} {\bibinfo {author} {\bibfnamefont {R.}~\bibnamefont
  {Arouca}}, \bibinfo {author} {\bibfnamefont {J.}~\bibnamefont {Cayao}},\ and\
  \bibinfo {author} {\bibfnamefont {A.~M.}\ \bibnamefont {Black-Schaffer}},\
  }\bibfield  {title} {\bibinfo {title} {Exceptionally enhanced topological
  superconductivity},\ }\href {https://arxiv.org/abs/2206.15324} {\bibfield
  {journal} {\bibinfo  {journal} {arXiv:2206.15324}\ } (\bibinfo {year}
  {2022})}\BibitemShut {NoStop}%
\bibitem [{\citenamefont {Hodaei}\ \emph {et~al.}(2017)\citenamefont {Hodaei},
  \citenamefont {Hassan}, \citenamefont {Wittek}, \citenamefont
  {Garcia-Gracia}, \citenamefont {El-Ganainy}, \citenamefont
  {Christodoulides},\ and\ \citenamefont {Khajavikhan}}]{hodaei2017enhanced}%
  \BibitemOpen
  \bibfield  {author} {\bibinfo {author} {\bibfnamefont {H.}~\bibnamefont
  {Hodaei}}, \bibinfo {author} {\bibfnamefont {A.~U.}\ \bibnamefont {Hassan}},
  \bibinfo {author} {\bibfnamefont {S.}~\bibnamefont {Wittek}}, \bibinfo
  {author} {\bibfnamefont {H.}~\bibnamefont {Garcia-Gracia}}, \bibinfo {author}
  {\bibfnamefont {R.}~\bibnamefont {El-Ganainy}}, \bibinfo {author}
  {\bibfnamefont {D.~N.}\ \bibnamefont {Christodoulides}},\ and\ \bibinfo
  {author} {\bibfnamefont {M.}~\bibnamefont {Khajavikhan}},\ }\bibfield
  {title} {\bibinfo {title} {Enhanced sensitivity at higher-order exceptional
  points},\ }\href {https://www.nature.com/articles/nature23280} {\bibfield
  {journal} {\bibinfo  {journal} {Nature}\ }\textbf {\bibinfo {volume} {548}},\
  \bibinfo {pages} {187} (\bibinfo {year} {2017})}\BibitemShut {NoStop}%
\bibitem [{\citenamefont {Chen}\ \emph {et~al.}(2017)\citenamefont {Chen},
  \citenamefont {{\"O}zdemir}, \citenamefont {Zhao}, \citenamefont {Wiersig},\
  and\ \citenamefont {Yang}}]{chen2017exceptional}%
  \BibitemOpen
  \bibfield  {author} {\bibinfo {author} {\bibfnamefont {W.}~\bibnamefont
  {Chen}}, \bibinfo {author} {\bibfnamefont {{\c{S}}.~K.}\ \bibnamefont
  {{\"O}zdemir}}, \bibinfo {author} {\bibfnamefont {G.}~\bibnamefont {Zhao}},
  \bibinfo {author} {\bibfnamefont {J.}~\bibnamefont {Wiersig}},\ and\ \bibinfo
  {author} {\bibfnamefont {L.}~\bibnamefont {Yang}},\ }\bibfield  {title}
  {\bibinfo {title} {Exceptional points enhance sensing in an optical
  microcavity},\ }\href {https://www.nature.com/articles/nature23281}
  {\bibfield  {journal} {\bibinfo  {journal} {Nature}\ }\textbf {\bibinfo
  {volume} {548}},\ \bibinfo {pages} {192} (\bibinfo {year}
  {2017})}\BibitemShut {NoStop}%
\bibitem [{\citenamefont {Peng}\ \emph {et~al.}(2016)\citenamefont {Peng},
  \citenamefont {{\"O}zdemir}, \citenamefont {Liertzer}, \citenamefont {Chen},
  \citenamefont {Kramer}, \citenamefont {Y{\i}lmaz}, \citenamefont {Wiersig},
  \citenamefont {Rotter},\ and\ \citenamefont {Yang}}]{peng2016chiral}%
  \BibitemOpen
  \bibfield  {author} {\bibinfo {author} {\bibfnamefont {B.}~\bibnamefont
  {Peng}}, \bibinfo {author} {\bibfnamefont {{\c{S}}.~K.}\ \bibnamefont
  {{\"O}zdemir}}, \bibinfo {author} {\bibfnamefont {M.}~\bibnamefont
  {Liertzer}}, \bibinfo {author} {\bibfnamefont {W.}~\bibnamefont {Chen}},
  \bibinfo {author} {\bibfnamefont {J.}~\bibnamefont {Kramer}}, \bibinfo
  {author} {\bibfnamefont {H.}~\bibnamefont {Y{\i}lmaz}}, \bibinfo {author}
  {\bibfnamefont {J.}~\bibnamefont {Wiersig}}, \bibinfo {author} {\bibfnamefont
  {S.}~\bibnamefont {Rotter}},\ and\ \bibinfo {author} {\bibfnamefont
  {L.}~\bibnamefont {Yang}},\ }\bibfield  {title} {\bibinfo {title} {Chiral
  modes and directional lasing at exceptional points},\ }\href
  {https://www.pnas.org/doi/10.1073/pnas.1603318113} {\bibfield  {journal}
  {\bibinfo  {journal} {Proc. Natl. Acad. Sci. U.S.A.}\ }\textbf {\bibinfo
  {volume} {113}},\ \bibinfo {pages} {6845} (\bibinfo {year}
  {2016})}\BibitemShut {NoStop}%
\bibitem [{\citenamefont {Longhi}\ and\ \citenamefont
  {Feng}(2017)}]{Longhi:17}%
  \BibitemOpen
  \bibfield  {author} {\bibinfo {author} {\bibfnamefont {S.}~\bibnamefont
  {Longhi}}\ and\ \bibinfo {author} {\bibfnamefont {L.}~\bibnamefont {Feng}},\
  }\bibfield  {title} {\bibinfo {title} {Unidirectional lasing in semiconductor
  microring lasers at an exceptional point},\ }\href
  {https://doi.org/10.1364/PRJ.5.0000B1} {\bibfield  {journal} {\bibinfo
  {journal} {Photon. Res.}\ }\textbf {\bibinfo {volume} {5}},\ \bibinfo {pages}
  {B1} (\bibinfo {year} {2017})}\BibitemShut {NoStop}%
\bibitem [{\citenamefont {Kozii}\ and\ \citenamefont
  {Fu}(2017)}]{kozii2017non}%
  \BibitemOpen
  \bibfield  {author} {\bibinfo {author} {\bibfnamefont {V.}~\bibnamefont
  {Kozii}}\ and\ \bibinfo {author} {\bibfnamefont {L.}~\bibnamefont {Fu}},\
  }\bibfield  {title} {\bibinfo {title} {Non-hermitian topological theory of
  finite-lifetime quasiparticles: prediction of bulk fermi arc due to
  exceptional point},\ }\href {https://arxiv.org/abs/1708.05841} {\bibfield
  {journal} {\bibinfo  {journal} {arXiv:1708.05841}\ } (\bibinfo {year}
  {2017})}\BibitemShut {NoStop}%
\bibitem [{\citenamefont {Ghatak}\ and\ \citenamefont
  {Das}(2018)}]{PhysRevB.97.014512}%
  \BibitemOpen
  \bibfield  {author} {\bibinfo {author} {\bibfnamefont {A.}~\bibnamefont
  {Ghatak}}\ and\ \bibinfo {author} {\bibfnamefont {T.}~\bibnamefont {Das}},\
  }\bibfield  {title} {\bibinfo {title} {Theory of superconductivity with
  non-hermitian and parity-time reversal symmetric cooper pairing symmetry},\
  }\href {https://doi.org/10.1103/PhysRevB.97.014512} {\bibfield  {journal}
  {\bibinfo  {journal} {Phys. Rev. B}\ }\textbf {\bibinfo {volume} {97}},\
  \bibinfo {pages} {014512} (\bibinfo {year} {2018})}\BibitemShut {NoStop}%
\bibitem [{\citenamefont {Yoshida}\ \emph {et~al.}(2018)\citenamefont
  {Yoshida}, \citenamefont {Peters},\ and\ \citenamefont
  {Kawakami}}]{PhysRevB.98.035141}%
  \BibitemOpen
  \bibfield  {author} {\bibinfo {author} {\bibfnamefont {T.}~\bibnamefont
  {Yoshida}}, \bibinfo {author} {\bibfnamefont {R.}~\bibnamefont {Peters}},\
  and\ \bibinfo {author} {\bibfnamefont {N.}~\bibnamefont {Kawakami}},\
  }\bibfield  {title} {\bibinfo {title} {Non-hermitian perspective of the band
  structure in heavy-fermion systems},\ }\href
  {https://doi.org/10.1103/PhysRevB.98.035141} {\bibfield  {journal} {\bibinfo
  {journal} {Phys. Rev. B}\ }\textbf {\bibinfo {volume} {98}},\ \bibinfo
  {pages} {035141} (\bibinfo {year} {2018})}\BibitemShut {NoStop}%
\bibitem [{\citenamefont {Okugawa}\ and\ \citenamefont
  {Yokoyama}(2019)}]{PhysRevB.99.041202}%
  \BibitemOpen
  \bibfield  {author} {\bibinfo {author} {\bibfnamefont {R.}~\bibnamefont
  {Okugawa}}\ and\ \bibinfo {author} {\bibfnamefont {T.}~\bibnamefont
  {Yokoyama}},\ }\bibfield  {title} {\bibinfo {title} {Topological exceptional
  surfaces in non-hermitian systems with parity-time and parity-particle-hole
  symmetries},\ }\href {https://doi.org/10.1103/PhysRevB.99.041202} {\bibfield
  {journal} {\bibinfo  {journal} {Phys. Rev. B}\ }\textbf {\bibinfo {volume}
  {99}},\ \bibinfo {pages} {041202} (\bibinfo {year} {2019})}\BibitemShut
  {NoStop}%
\bibitem [{\citenamefont {Kawabata}\ \emph
  {et~al.}(2019{\natexlab{b}})\citenamefont {Kawabata}, \citenamefont
  {Bessho},\ and\ \citenamefont {Sato}}]{PhysRevLett.123.066405}%
  \BibitemOpen
  \bibfield  {author} {\bibinfo {author} {\bibfnamefont {K.}~\bibnamefont
  {Kawabata}}, \bibinfo {author} {\bibfnamefont {T.}~\bibnamefont {Bessho}},\
  and\ \bibinfo {author} {\bibfnamefont {M.}~\bibnamefont {Sato}},\ }\bibfield
  {title} {\bibinfo {title} {Classification of exceptional points and
  non-hermitian topological semimetals},\ }\href
  {https://doi.org/10.1103/PhysRevLett.123.066405} {\bibfield  {journal}
  {\bibinfo  {journal} {Phys. Rev. Lett.}\ }\textbf {\bibinfo {volume} {123}},\
  \bibinfo {pages} {066405} (\bibinfo {year} {2019}{\natexlab{b}})}\BibitemShut
  {NoStop}%
\bibitem [{\citenamefont {Yamamoto}\ \emph {et~al.}(2019)\citenamefont
  {Yamamoto}, \citenamefont {Nakagawa}, \citenamefont {Adachi}, \citenamefont
  {Takasan}, \citenamefont {Ueda},\ and\ \citenamefont
  {Kawakami}}]{PhysRevLett.123.123601}%
  \BibitemOpen
  \bibfield  {author} {\bibinfo {author} {\bibfnamefont {K.}~\bibnamefont
  {Yamamoto}}, \bibinfo {author} {\bibfnamefont {M.}~\bibnamefont {Nakagawa}},
  \bibinfo {author} {\bibfnamefont {K.}~\bibnamefont {Adachi}}, \bibinfo
  {author} {\bibfnamefont {K.}~\bibnamefont {Takasan}}, \bibinfo {author}
  {\bibfnamefont {M.}~\bibnamefont {Ueda}},\ and\ \bibinfo {author}
  {\bibfnamefont {N.}~\bibnamefont {Kawakami}},\ }\bibfield  {title} {\bibinfo
  {title} {Theory of non-hermitian fermionic superfluidity with a
  complex-valued interaction},\ }\href
  {https://doi.org/10.1103/PhysRevLett.123.123601} {\bibfield  {journal}
  {\bibinfo  {journal} {Phys. Rev. Lett.}\ }\textbf {\bibinfo {volume} {123}},\
  \bibinfo {pages} {123601} (\bibinfo {year} {2019})}\BibitemShut {NoStop}%
\bibitem [{\citenamefont {Zhou}\ \emph {et~al.}(2018)\citenamefont {Zhou},
  \citenamefont {Peng}, \citenamefont {Yoon}, \citenamefont {Hsu},
  \citenamefont {Nelson}, \citenamefont {Fu}, \citenamefont {Joannopoulos},
  \citenamefont {Solja\v{c}i\'{c}},\ and\ \citenamefont
  {Zhen}}]{science359Zhou}%
  \BibitemOpen
  \bibfield  {author} {\bibinfo {author} {\bibfnamefont {H.}~\bibnamefont
  {Zhou}}, \bibinfo {author} {\bibfnamefont {C.}~\bibnamefont {Peng}}, \bibinfo
  {author} {\bibfnamefont {Y.}~\bibnamefont {Yoon}}, \bibinfo {author}
  {\bibfnamefont {C.~W.}\ \bibnamefont {Hsu}}, \bibinfo {author} {\bibfnamefont
  {K.~A.}\ \bibnamefont {Nelson}}, \bibinfo {author} {\bibfnamefont
  {L.}~\bibnamefont {Fu}}, \bibinfo {author} {\bibfnamefont {J.~D.}\
  \bibnamefont {Joannopoulos}}, \bibinfo {author} {\bibfnamefont
  {M.}~\bibnamefont {Solja\v{c}i\'{c}}},\ and\ \bibinfo {author} {\bibfnamefont
  {B.}~\bibnamefont {Zhen}},\ }\bibfield  {title} {\bibinfo {title}
  {Observation of bulk fermi arc and polarization half charge from paired
  exceptional points},\ }\href {https://doi.org/10.1126/science.aap9859}
  {\bibfield  {journal} {\bibinfo  {journal} {Science}\ }\textbf {\bibinfo
  {volume} {359}},\ \bibinfo {pages} {1009} (\bibinfo {year}
  {2018})}\BibitemShut {NoStop}%
\bibitem [{\citenamefont {Bessho}\ \emph {et~al.}(2019)\citenamefont {Bessho},
  \citenamefont {Kawabata},\ and\ \citenamefont
  {Sato}}]{doi:10.7566/JPSCP.30.011098}%
  \BibitemOpen
  \bibfield  {author} {\bibinfo {author} {\bibfnamefont {T.}~\bibnamefont
  {Bessho}}, \bibinfo {author} {\bibfnamefont {K.}~\bibnamefont {Kawabata}},\
  and\ \bibinfo {author} {\bibfnamefont {M.}~\bibnamefont {Sato}},\ }\bibinfo
  {title} {Topological classificaton of non-hermitian gapless phases:
  Exceptional points and bulk fermi arcs},\ in\ \href
  {https://doi.org/10.7566/JPSCP.30.011098} {\emph {\bibinfo {booktitle} {Proc.
  Int. Conf. on Strongly Correlated Electron Systems (SCES2019)}}}\ (\bibinfo
  {publisher} {Physical Society of Japan},\ \bibinfo {year} {2019})\
  Chap.~\bibinfo {chapter} {30}, p.\ \bibinfo {pages} {011098}\BibitemShut
  {NoStop}%
\bibitem [{\citenamefont {Nagai}\ \emph {et~al.}(2020)\citenamefont {Nagai},
  \citenamefont {Qi}, \citenamefont {Isobe}, \citenamefont {Kozii},\ and\
  \citenamefont {Fu}}]{PhysRevLett.125.227204}%
  \BibitemOpen
  \bibfield  {author} {\bibinfo {author} {\bibfnamefont {Y.}~\bibnamefont
  {Nagai}}, \bibinfo {author} {\bibfnamefont {Y.}~\bibnamefont {Qi}}, \bibinfo
  {author} {\bibfnamefont {H.}~\bibnamefont {Isobe}}, \bibinfo {author}
  {\bibfnamefont {V.}~\bibnamefont {Kozii}},\ and\ \bibinfo {author}
  {\bibfnamefont {L.}~\bibnamefont {Fu}},\ }\bibfield  {title} {\bibinfo
  {title} {Dmft reveals the non-hermitian topology and fermi arcs in
  heavy-fermion systems},\ }\href
  {https://doi.org/10.1103/PhysRevLett.125.227204} {\bibfield  {journal}
  {\bibinfo  {journal} {Phys. Rev. Lett.}\ }\textbf {\bibinfo {volume} {125}},\
  \bibinfo {pages} {227204} (\bibinfo {year} {2020})}\BibitemShut {NoStop}%
\bibitem [{\citenamefont {Mandal}\ and\ \citenamefont
  {Bergholtz}(2021)}]{PhysRevLett.127.186601}%
  \BibitemOpen
  \bibfield  {author} {\bibinfo {author} {\bibfnamefont {I.}~\bibnamefont
  {Mandal}}\ and\ \bibinfo {author} {\bibfnamefont {E.~J.}\ \bibnamefont
  {Bergholtz}},\ }\bibfield  {title} {\bibinfo {title} {Symmetry and
  higher-order exceptional points},\ }\href
  {https://doi.org/10.1103/PhysRevLett.127.186601} {\bibfield  {journal}
  {\bibinfo  {journal} {Phys. Rev. Lett.}\ }\textbf {\bibinfo {volume} {127}},\
  \bibinfo {pages} {186601} (\bibinfo {year} {2021})}\BibitemShut {NoStop}%
\bibitem [{\citenamefont {Delplace}\ \emph {et~al.}(2021)\citenamefont
  {Delplace}, \citenamefont {Yoshida},\ and\ \citenamefont
  {Hatsugai}}]{PhysRevLett.127.186602}%
  \BibitemOpen
  \bibfield  {author} {\bibinfo {author} {\bibfnamefont {P.}~\bibnamefont
  {Delplace}}, \bibinfo {author} {\bibfnamefont {T.}~\bibnamefont {Yoshida}},\
  and\ \bibinfo {author} {\bibfnamefont {Y.}~\bibnamefont {Hatsugai}},\
  }\bibfield  {title} {\bibinfo {title} {Symmetry-protected multifold
  exceptional points and their topological characterization},\ }\href
  {https://doi.org/10.1103/PhysRevLett.127.186602} {\bibfield  {journal}
  {\bibinfo  {journal} {Phys. Rev. Lett.}\ }\textbf {\bibinfo {volume} {127}},\
  \bibinfo {pages} {186602} (\bibinfo {year} {2021})}\BibitemShut {NoStop}%
\bibitem [{\citenamefont {Yoshida}\ \emph {et~al.}(2020)\citenamefont
  {Yoshida}, \citenamefont {Peters}, \citenamefont {Kawakami},\ and\
  \citenamefont {Hatsugai}}]{yoshida2020exceptional}%
  \BibitemOpen
  \bibfield  {author} {\bibinfo {author} {\bibfnamefont {T.}~\bibnamefont
  {Yoshida}}, \bibinfo {author} {\bibfnamefont {R.}~\bibnamefont {Peters}},
  \bibinfo {author} {\bibfnamefont {N.}~\bibnamefont {Kawakami}},\ and\
  \bibinfo {author} {\bibfnamefont {Y.}~\bibnamefont {Hatsugai}},\ }\bibfield
  {title} {\bibinfo {title} {Exceptional band touching for strongly correlated
  systems in equilibrium},\ }\href
  {https://academic.oup.com/ptep/article/2020/12/12A109/5875995} {\bibfield
  {journal} {\bibinfo  {journal} {Prog. Theor. Exp. Phys.}\ }\textbf {\bibinfo
  {volume} {2020}},\ \bibinfo {pages} {12A109} (\bibinfo {year}
  {2020})}\BibitemShut {NoStop}%
\bibitem [{\citenamefont {Rausch}\ \emph {et~al.}(2021)\citenamefont {Rausch},
  \citenamefont {Peters},\ and\ \citenamefont
  {Yoshida}}]{rausch2021exceptional}%
  \BibitemOpen
  \bibfield  {author} {\bibinfo {author} {\bibfnamefont {R.}~\bibnamefont
  {Rausch}}, \bibinfo {author} {\bibfnamefont {R.}~\bibnamefont {Peters}},\
  and\ \bibinfo {author} {\bibfnamefont {T.}~\bibnamefont {Yoshida}},\
  }\bibfield  {title} {\bibinfo {title} {Exceptional points in the
  one-dimensional hubbard model},\ }\href
  {https://iopscience.iop.org/article/10.1088/1367-2630/abd35e} {\bibfield
  {journal} {\bibinfo  {journal} {New J. Phys.}\ }\textbf {\bibinfo {volume}
  {23}},\ \bibinfo {pages} {013011} (\bibinfo {year} {2021})}\BibitemShut
  {NoStop}%
\bibitem [{\citenamefont {Pikulin}\ and\ \citenamefont
  {Nazarov}(2012)}]{pikulin2012topological}%
  \BibitemOpen
  \bibfield  {author} {\bibinfo {author} {\bibfnamefont {D.}~\bibnamefont
  {Pikulin}}\ and\ \bibinfo {author} {\bibfnamefont {Y.~V.}\ \bibnamefont
  {Nazarov}},\ }\bibfield  {title} {\bibinfo {title} {Topological properties of
  superconducting junctions},\ }\href
  {https://link.springer.com/article/10.1134/S0021364011210090} {\bibfield
  {journal} {\bibinfo  {journal} {JETP letters}\ }\textbf {\bibinfo {volume}
  {94}},\ \bibinfo {pages} {693} (\bibinfo {year} {2012})}\BibitemShut
  {NoStop}%
\bibitem [{\citenamefont {Pikulin}\ and\ \citenamefont
  {Nazarov}(2013)}]{pikulin2013two}%
  \BibitemOpen
  \bibfield  {author} {\bibinfo {author} {\bibfnamefont {D.~I.}\ \bibnamefont
  {Pikulin}}\ and\ \bibinfo {author} {\bibfnamefont {Y.~V.}\ \bibnamefont
  {Nazarov}},\ }\bibfield  {title} {\bibinfo {title} {Two types of topological
  transitions in finite majorana wires},\ }\href
  {https://link.aps.org/doi/10.1103/PhysRevB.87.235421} {\bibfield  {journal}
  {\bibinfo  {journal} {Phys. Rev. B}\ }\textbf {\bibinfo {volume} {87}},\
  \bibinfo {pages} {235421} (\bibinfo {year} {2013})}\BibitemShut {NoStop}%
\bibitem [{\citenamefont {San-Jose}\ \emph {et~al.}(2016)\citenamefont
  {San-Jose}, \citenamefont {Cayao}, \citenamefont {Prada},\ and\ \citenamefont
  {Aguado}}]{san2016majorana}%
  \BibitemOpen
  \bibfield  {author} {\bibinfo {author} {\bibfnamefont {P.}~\bibnamefont
  {San-Jose}}, \bibinfo {author} {\bibfnamefont {J.}~\bibnamefont {Cayao}},
  \bibinfo {author} {\bibfnamefont {E.}~\bibnamefont {Prada}},\ and\ \bibinfo
  {author} {\bibfnamefont {R.}~\bibnamefont {Aguado}},\ }\bibfield  {title}
  {\bibinfo {title} {Majorana bound states from exceptional points in
  non-topological superconductors},\ }\href
  {https://www.nature.com/articles/srep21427} {\bibfield  {journal} {\bibinfo
  {journal} {Sci. Rep.}\ }\textbf {\bibinfo {volume} {6}},\ \bibinfo {pages}
  {21427} (\bibinfo {year} {2016})}\BibitemShut {NoStop}%
\bibitem [{\citenamefont {Avila}\ \emph {et~al.}(2019)\citenamefont {Avila},
  \citenamefont {Pe{\~n}aranda}, \citenamefont {Prada}, \citenamefont
  {San-Jose},\ and\ \citenamefont {Aguado}}]{avila2019non}%
  \BibitemOpen
  \bibfield  {author} {\bibinfo {author} {\bibfnamefont {J.}~\bibnamefont
  {Avila}}, \bibinfo {author} {\bibfnamefont {F.}~\bibnamefont
  {Pe{\~n}aranda}}, \bibinfo {author} {\bibfnamefont {E.}~\bibnamefont
  {Prada}}, \bibinfo {author} {\bibfnamefont {P.}~\bibnamefont {San-Jose}},\
  and\ \bibinfo {author} {\bibfnamefont {R.}~\bibnamefont {Aguado}},\
  }\bibfield  {title} {\bibinfo {title} {Non-hermitian topology as a unifying
  framework for the andreev versus majorana states controversy},\ }\href
  {https://www.nature.com/articles/s42005-019-0231-8} {\bibfield  {journal}
  {\bibinfo  {journal} {Commun. Phys.}\ }\textbf {\bibinfo {volume} {2}},\
  \bibinfo {pages} {133} (\bibinfo {year} {2019})}\BibitemShut {NoStop}%
\bibitem [{\citenamefont {Philip}\ \emph {et~al.}(2018)\citenamefont {Philip},
  \citenamefont {Hirsbrunner},\ and\ \citenamefont
  {Gilbert}}]{PhysRevB.98.155430}%
  \BibitemOpen
  \bibfield  {author} {\bibinfo {author} {\bibfnamefont {T.~M.}\ \bibnamefont
  {Philip}}, \bibinfo {author} {\bibfnamefont {M.~R.}\ \bibnamefont
  {Hirsbrunner}},\ and\ \bibinfo {author} {\bibfnamefont {M.~J.}\ \bibnamefont
  {Gilbert}},\ }\bibfield  {title} {\bibinfo {title} {Loss of hall conductivity
  quantization in a non-hermitian quantum anomalous hall insulator},\ }\href
  {https://doi.org/10.1103/PhysRevB.98.155430} {\bibfield  {journal} {\bibinfo
  {journal} {Phys. Rev. B}\ }\textbf {\bibinfo {volume} {98}},\ \bibinfo
  {pages} {155430} (\bibinfo {year} {2018})}\BibitemShut {NoStop}%
\bibitem [{\citenamefont {Chen}\ and\ \citenamefont
  {Zhai}(2018)}]{PhysRevB.98.245130}%
  \BibitemOpen
  \bibfield  {author} {\bibinfo {author} {\bibfnamefont {Y.}~\bibnamefont
  {Chen}}\ and\ \bibinfo {author} {\bibfnamefont {H.}~\bibnamefont {Zhai}},\
  }\bibfield  {title} {\bibinfo {title} {Hall conductance of a non-hermitian
  chern insulator},\ }\href {https://doi.org/10.1103/PhysRevB.98.245130}
  {\bibfield  {journal} {\bibinfo  {journal} {Phys. Rev. B}\ }\textbf {\bibinfo
  {volume} {98}},\ \bibinfo {pages} {245130} (\bibinfo {year}
  {2018})}\BibitemShut {NoStop}%
\bibitem [{\citenamefont {Bergholtz}\ and\ \citenamefont
  {Budich}(2019)}]{PhysRevResearch.1.012003}%
  \BibitemOpen
  \bibfield  {author} {\bibinfo {author} {\bibfnamefont {E.~J.}\ \bibnamefont
  {Bergholtz}}\ and\ \bibinfo {author} {\bibfnamefont {J.~C.}\ \bibnamefont
  {Budich}},\ }\bibfield  {title} {\bibinfo {title} {Non-hermitian weyl physics
  in topological insulator ferromagnet junctions},\ }\href
  {https://doi.org/10.1103/PhysRevResearch.1.012003} {\bibfield  {journal}
  {\bibinfo  {journal} {Phys. Rev. Research}\ }\textbf {\bibinfo {volume}
  {1}},\ \bibinfo {pages} {012003} (\bibinfo {year} {2019})}\BibitemShut
  {NoStop}%
\bibitem [{\citenamefont {Cayao}\ and\ \citenamefont
  {Black-Schaffer}(2022{\natexlab{a}})}]{PhysRevB.105.094502}%
  \BibitemOpen
  \bibfield  {author} {\bibinfo {author} {\bibfnamefont {J.}~\bibnamefont
  {Cayao}}\ and\ \bibinfo {author} {\bibfnamefont {A.~M.}\ \bibnamefont
  {Black-Schaffer}},\ }\bibfield  {title} {\bibinfo {title} {Exceptional
  odd-frequency pairing in non-hermitian superconducting systems},\ }\href
  {https://doi.org/10.1103/PhysRevB.105.094502} {\bibfield  {journal} {\bibinfo
   {journal} {Phys. Rev. B}\ }\textbf {\bibinfo {volume} {105}},\ \bibinfo
  {pages} {094502} (\bibinfo {year} {2022}{\natexlab{a}})}\BibitemShut
  {NoStop}%
\bibitem [{\citenamefont {Cayao}\ and\ \citenamefont
  {Black-Schaffer}(2022{\natexlab{b}})}]{cayao2022NH}%
  \BibitemOpen
  \bibfield  {author} {\bibinfo {author} {\bibfnamefont {J.}~\bibnamefont
  {Cayao}}\ and\ \bibinfo {author} {\bibfnamefont {A.~M.}\ \bibnamefont
  {Black-Schaffer}},\ }\bibfield  {title} {\bibinfo {title} {Bulk bogoliubov
  fermi arcs in non-hermitian superconducting systems},\ }\href
  {https://arxiv.org/abs/2208.05372} {\bibfield  {journal} {\bibinfo  {journal}
  {arxiv.2208.05372}\ } (\bibinfo {year} {2022}{\natexlab{b}})}\BibitemShut
  {NoStop}%
\bibitem [{\citenamefont {Datta}(1997)}]{datta1997electronic}%
  \BibitemOpen
  \bibfield  {author} {\bibinfo {author} {\bibfnamefont {S.}~\bibnamefont
  {Datta}},\ }\href@noop {} {\emph {\bibinfo {title} {Electronic transport in
  mesoscopic systems}}}\ (\bibinfo  {publisher} {Cambridge university press},\
  \bibinfo {year} {1997})\BibitemShut {NoStop}%
\bibitem [{\citenamefont {Dresselhaus}(1955)}]{PhysRev.100.580}%
  \BibitemOpen
  \bibfield  {author} {\bibinfo {author} {\bibfnamefont {G.}~\bibnamefont
  {Dresselhaus}},\ }\bibfield  {title} {\bibinfo {title} {Spin-orbit coupling
  effects in zinc blende structures},\ }\href
  {https://doi.org/10.1103/PhysRev.100.580} {\bibfield  {journal} {\bibinfo
  {journal} {Phys. Rev.}\ }\textbf {\bibinfo {volume} {100}},\ \bibinfo {pages}
  {580} (\bibinfo {year} {1955})}\BibitemShut {NoStop}%
\bibitem [{\citenamefont {Rashba}(1960)}]{Rashba1960}%
  \BibitemOpen
  \bibfield  {author} {\bibinfo {author} {\bibfnamefont {E.}~\bibnamefont
  {Rashba}},\ }\bibfield  {title} {\bibinfo {title} {Properties of
  semiconductors with an extremum loop},\ }\href@noop {} {\bibfield  {journal}
  {\bibinfo  {journal} {Sov. Phys. Solid. State}\ }\textbf {\bibinfo {volume}
  {2}},\ \bibinfo {pages} {1109} (\bibinfo {year} {1960})}\BibitemShut
  {NoStop}%
\bibitem [{\citenamefont {Bychkov}\ and\ \citenamefont
  {Rashba}(1984)}]{rashba84a}%
  \BibitemOpen
  \bibfield  {author} {\bibinfo {author} {\bibfnamefont {Y.~A.}\ \bibnamefont
  {Bychkov}}\ and\ \bibinfo {author} {\bibfnamefont {E.~I.}\ \bibnamefont
  {Rashba}},\ }\bibfield  {title} {\bibinfo {title} {Properties of a 2d
  electron gas with lifted spectral degeneracy},\ }\href
  {http://jetpletters.ru/ps/1264/article_19121.shtml} {\bibfield  {journal}
  {\bibinfo  {journal} {Sov. Phys. JETP}\ }\textbf {\bibinfo {volume} {39}},\
  \bibinfo {pages} {78} (\bibinfo {year} {1984})}\BibitemShut {NoStop}%
\bibitem [{\citenamefont {Winkler}(2003)}]{winkler2003spin}%
  \BibitemOpen
  \bibfield  {author} {\bibinfo {author} {\bibfnamefont {R.}~\bibnamefont
  {Winkler}},\ }\href {https://link.springer.com/book/10.1007/b13586} {\emph
  {\bibinfo {title} {Spin-orbit coupling effects in two-dimensional electron
  and hole systems}}},\ Vol.\ \bibinfo {volume} {191}\ (\bibinfo  {publisher}
  {Springer},\ \bibinfo {year} {2003})\BibitemShut {NoStop}%
\bibitem [{\citenamefont {Chen}\ \emph
  {et~al.}(2021{\natexlab{a}})\citenamefont {Chen}, \citenamefont {Wu},
  \citenamefont {Hu},\ and\ \citenamefont {Yang}}]{chen2021spin}%
  \BibitemOpen
  \bibfield  {author} {\bibinfo {author} {\bibfnamefont {J.}~\bibnamefont
  {Chen}}, \bibinfo {author} {\bibfnamefont {K.}~\bibnamefont {Wu}}, \bibinfo
  {author} {\bibfnamefont {W.}~\bibnamefont {Hu}},\ and\ \bibinfo {author}
  {\bibfnamefont {J.}~\bibnamefont {Yang}},\ }\bibfield  {title} {\bibinfo
  {title} {Spin--orbit coupling in 2d semiconductors: A theoretical
  perspective},\ }\href {https://pubs.acs.org/doi/10.1021/acs.jpclett.1c03662}
  {\bibfield  {journal} {\bibinfo  {journal} {J. Phys. Chem. Lett.}\ }\textbf
  {\bibinfo {volume} {12}},\ \bibinfo {pages} {12256} (\bibinfo {year}
  {2021}{\natexlab{a}})}\BibitemShut {NoStop}%
\bibitem [{\citenamefont {Manchon}\ \emph {et~al.}(2015)\citenamefont
  {Manchon}, \citenamefont {Koo}, \citenamefont {Nitta}, \citenamefont
  {Frolov},\ and\ \citenamefont {Duine}}]{manchon2015new}%
  \BibitemOpen
  \bibfield  {author} {\bibinfo {author} {\bibfnamefont {A.}~\bibnamefont
  {Manchon}}, \bibinfo {author} {\bibfnamefont {H.~C.}\ \bibnamefont {Koo}},
  \bibinfo {author} {\bibfnamefont {J.}~\bibnamefont {Nitta}}, \bibinfo
  {author} {\bibfnamefont {S.~M.}\ \bibnamefont {Frolov}},\ and\ \bibinfo
  {author} {\bibfnamefont {R.~A.}\ \bibnamefont {Duine}},\ }\bibfield  {title}
  {\bibinfo {title} {New perspectives for rashba spin--orbit coupling},\ }\href
  {https://www.nature.com/articles/nmat4360} {\bibfield  {journal} {\bibinfo
  {journal} {Nat. Mater.}\ }\textbf {\bibinfo {volume} {14}},\ \bibinfo {pages}
  {871} (\bibinfo {year} {2015})}\BibitemShut {NoStop}%
\bibitem [{\citenamefont {Lutchyn}\ \emph {et~al.}(2018)\citenamefont
  {Lutchyn}, \citenamefont {Bakkers}, \citenamefont {Kouwenhoven},
  \citenamefont {Krogstrup}, \citenamefont {Marcus},\ and\ \citenamefont
  {Oreg}}]{Lutchyn2018Majorana}%
  \BibitemOpen
  \bibfield  {author} {\bibinfo {author} {\bibfnamefont {R.~M.}\ \bibnamefont
  {Lutchyn}}, \bibinfo {author} {\bibfnamefont {E.~P. A.~M.}\ \bibnamefont
  {Bakkers}}, \bibinfo {author} {\bibfnamefont {L.~P.}\ \bibnamefont
  {Kouwenhoven}}, \bibinfo {author} {\bibfnamefont {P.}~\bibnamefont
  {Krogstrup}}, \bibinfo {author} {\bibfnamefont {C.~M.}\ \bibnamefont
  {Marcus}},\ and\ \bibinfo {author} {\bibfnamefont {Y.}~\bibnamefont {Oreg}},\
  }\bibfield  {title} {\bibinfo {title} {Majorana zero modes in
  superconductor-semiconductor heterostructures},\ }\href
  {https://doi.org/10.1038/s41578-018-0003-1} {\bibfield  {journal} {\bibinfo
  {journal} {Nat. Rev. Mater.}\ }\textbf {\bibinfo {volume} {3}},\ \bibinfo
  {pages} {52} (\bibinfo {year} {2018})}\BibitemShut {NoStop}%
\bibitem [{\citenamefont {Prada}\ \emph {et~al.}(2020)\citenamefont {Prada},
  \citenamefont {San-Jose}, \citenamefont {de~Moor}, \citenamefont {Geresdi},
  \citenamefont {Lee}, \citenamefont {Klinovaja}, \citenamefont {Loss},
  \citenamefont {Nyg{\aa}rd}, \citenamefont {Aguado},\ and\ \citenamefont
  {Kouwenhoven}}]{prada2019andreev}%
  \BibitemOpen
  \bibfield  {author} {\bibinfo {author} {\bibfnamefont {E.}~\bibnamefont
  {Prada}}, \bibinfo {author} {\bibfnamefont {P.}~\bibnamefont {San-Jose}},
  \bibinfo {author} {\bibfnamefont {M.~W.}\ \bibnamefont {de~Moor}}, \bibinfo
  {author} {\bibfnamefont {A.}~\bibnamefont {Geresdi}}, \bibinfo {author}
  {\bibfnamefont {E.~J.}\ \bibnamefont {Lee}}, \bibinfo {author} {\bibfnamefont
  {J.}~\bibnamefont {Klinovaja}}, \bibinfo {author} {\bibfnamefont
  {D.}~\bibnamefont {Loss}}, \bibinfo {author} {\bibfnamefont {J.}~\bibnamefont
  {Nyg{\aa}rd}}, \bibinfo {author} {\bibfnamefont {R.}~\bibnamefont {Aguado}},\
  and\ \bibinfo {author} {\bibfnamefont {L.~P.}\ \bibnamefont {Kouwenhoven}},\
  }\bibfield  {title} {\bibinfo {title} {From {A}ndreev to {M}ajorana bound
  states in hybrid superconductor--semiconductor nanowires},\ }\href
  {https://doi.org/10.1038/s42254-020-0228-y} {\bibfield  {journal} {\bibinfo
  {journal} {Nat. Rev. Phys.}\ }\textbf {\bibinfo {volume} {2}},\ \bibinfo
  {pages} {575} (\bibinfo {year} {2020})}\BibitemShut {NoStop}%
\bibitem [{\citenamefont {Flensberg}\ \emph {et~al.}(2021)\citenamefont
  {Flensberg}, \citenamefont {von Oppen},\ and\ \citenamefont
  {Stern}}]{flensberg2021engineered}%
  \BibitemOpen
  \bibfield  {author} {\bibinfo {author} {\bibfnamefont {K.}~\bibnamefont
  {Flensberg}}, \bibinfo {author} {\bibfnamefont {F.}~\bibnamefont {von
  Oppen}},\ and\ \bibinfo {author} {\bibfnamefont {A.}~\bibnamefont {Stern}},\
  }\bibfield  {title} {\bibinfo {title} {Engineered platforms for topological
  superconductivity and majorana zero modes},\ }\href
  {https://www.nature.com/articles/s41578-021-00336-6} {\bibfield  {journal}
  {\bibinfo  {journal} {Nat. Rev. Mater.}\ }\textbf {\bibinfo {volume} {6}},\
  \bibinfo {pages} {944} (\bibinfo {year} {2021})}\BibitemShut {NoStop}%
\bibitem [{\citenamefont {Sestoft}\ \emph {et~al.}(2018)\citenamefont
  {Sestoft}, \citenamefont {Kanne}, \citenamefont {Gejl}, \citenamefont {von
  Soosten}, \citenamefont {Yodh}, \citenamefont {Sherman}, \citenamefont
  {Tarasinski}, \citenamefont {Wimmer}, \citenamefont {Johnson}, \citenamefont
  {Deng}, \citenamefont {Nyg\aa{}rd}, \citenamefont {Jespersen}, \citenamefont
  {Marcus},\ and\ \citenamefont {Krogstrup}}]{PhysRevMaterials.2.044202}%
  \BibitemOpen
  \bibfield  {author} {\bibinfo {author} {\bibfnamefont {J.~E.}\ \bibnamefont
  {Sestoft}}, \bibinfo {author} {\bibfnamefont {T.}~\bibnamefont {Kanne}},
  \bibinfo {author} {\bibfnamefont {A.~N.}\ \bibnamefont {Gejl}}, \bibinfo
  {author} {\bibfnamefont {M.}~\bibnamefont {von Soosten}}, \bibinfo {author}
  {\bibfnamefont {J.~S.}\ \bibnamefont {Yodh}}, \bibinfo {author}
  {\bibfnamefont {D.}~\bibnamefont {Sherman}}, \bibinfo {author} {\bibfnamefont
  {B.}~\bibnamefont {Tarasinski}}, \bibinfo {author} {\bibfnamefont
  {M.}~\bibnamefont {Wimmer}}, \bibinfo {author} {\bibfnamefont
  {E.}~\bibnamefont {Johnson}}, \bibinfo {author} {\bibfnamefont
  {M.}~\bibnamefont {Deng}}, \bibinfo {author} {\bibfnamefont {J.}~\bibnamefont
  {Nyg\aa{}rd}}, \bibinfo {author} {\bibfnamefont {T.~S.}\ \bibnamefont
  {Jespersen}}, \bibinfo {author} {\bibfnamefont {C.~M.}\ \bibnamefont
  {Marcus}},\ and\ \bibinfo {author} {\bibfnamefont {P.}~\bibnamefont
  {Krogstrup}},\ }\bibfield  {title} {\bibinfo {title} {Engineering hybrid
  epitaxial inassb/al nanowires for stronger topological protection},\ }\href
  {https://doi.org/10.1103/PhysRevMaterials.2.044202} {\bibfield  {journal}
  {\bibinfo  {journal} {Phys. Rev. Materials}\ }\textbf {\bibinfo {volume}
  {2}},\ \bibinfo {pages} {044202} (\bibinfo {year} {2018})}\BibitemShut
  {NoStop}%
\bibitem [{\citenamefont {Kammhuber}\ \emph {et~al.}(2017)\citenamefont
  {Kammhuber}, \citenamefont {Cassidy}, \citenamefont {Pei}, \citenamefont
  {Nowak}, \citenamefont {Vuik}, \citenamefont {G{\"u}l}, \citenamefont {Car},
  \citenamefont {Plissard}, \citenamefont {Bakkers}, \citenamefont {Wimmer}
  \emph {et~al.}}]{kammhuber2017conductance}%
  \BibitemOpen
  \bibfield  {author} {\bibinfo {author} {\bibfnamefont {J.}~\bibnamefont
  {Kammhuber}}, \bibinfo {author} {\bibfnamefont {M.~C.}\ \bibnamefont
  {Cassidy}}, \bibinfo {author} {\bibfnamefont {F.}~\bibnamefont {Pei}},
  \bibinfo {author} {\bibfnamefont {M.~P.}\ \bibnamefont {Nowak}}, \bibinfo
  {author} {\bibfnamefont {A.}~\bibnamefont {Vuik}}, \bibinfo {author}
  {\bibfnamefont {{\"O}.}~\bibnamefont {G{\"u}l}}, \bibinfo {author}
  {\bibfnamefont {D.}~\bibnamefont {Car}}, \bibinfo {author} {\bibfnamefont
  {S.}~\bibnamefont {Plissard}}, \bibinfo {author} {\bibfnamefont
  {E.}~\bibnamefont {Bakkers}}, \bibinfo {author} {\bibfnamefont
  {M.}~\bibnamefont {Wimmer}}, \emph {et~al.},\ }\bibfield  {title} {\bibinfo
  {title} {Conductance through a helical state in an indium antimonide
  nanowire},\ }\href {https://www.nature.com/articles/s41467-017-00315-y}
  {\bibfield  {journal} {\bibinfo  {journal} {Nat. Commun.}\ }\textbf {\bibinfo
  {volume} {8}},\ \bibinfo {pages} {478} (\bibinfo {year} {2017})}\BibitemShut
  {NoStop}%
\bibitem [{\citenamefont {Hu}\ \emph {et~al.}(2022)\citenamefont {Hu},
  \citenamefont {Sun},\ and\ \citenamefont {Chen}}]{PhysRevResearch.4.L022064}%
  \BibitemOpen
  \bibfield  {author} {\bibinfo {author} {\bibfnamefont {H.}~\bibnamefont
  {Hu}}, \bibinfo {author} {\bibfnamefont {S.}~\bibnamefont {Sun}},\ and\
  \bibinfo {author} {\bibfnamefont {S.}~\bibnamefont {Chen}},\ }\bibfield
  {title} {\bibinfo {title} {Knot topology of exceptional point and
  non-hermitian no-go theorem},\ }\href
  {https://doi.org/10.1103/PhysRevResearch.4.L022064} {\bibfield  {journal}
  {\bibinfo  {journal} {Phys. Rev. Res.}\ }\textbf {\bibinfo {volume} {4}},\
  \bibinfo {pages} {L022064} (\bibinfo {year} {2022})}\BibitemShut {NoStop}%
\bibitem [{\citenamefont {Cayao}\ \emph {et~al.}(2015)\citenamefont {Cayao},
  \citenamefont {Prada}, \citenamefont {San-Jose},\ and\ \citenamefont
  {Aguado}}]{PhysRevB.91.024514}%
  \BibitemOpen
  \bibfield  {author} {\bibinfo {author} {\bibfnamefont {J.}~\bibnamefont
  {Cayao}}, \bibinfo {author} {\bibfnamefont {E.}~\bibnamefont {Prada}},
  \bibinfo {author} {\bibfnamefont {P.}~\bibnamefont {San-Jose}},\ and\
  \bibinfo {author} {\bibfnamefont {R.}~\bibnamefont {Aguado}},\ }\bibfield
  {title} {\bibinfo {title} {Sns junctions in nanowires with spin-orbit
  coupling: Role of confinement and helicity on the subgap spectrum},\ }\href
  {https://doi.org/10.1103/PhysRevB.91.024514} {\bibfield  {journal} {\bibinfo
  {journal} {Phys. Rev. B}\ }\textbf {\bibinfo {volume} {91}},\ \bibinfo
  {pages} {024514} (\bibinfo {year} {2015})}\BibitemShut {NoStop}%
\bibitem [{\citenamefont {Oshima}\ \emph {et~al.}(2018)\citenamefont {Oshima},
  \citenamefont {Taguchi},\ and\ \citenamefont {Tanaka}}]{oshima2018tunneling}%
  \BibitemOpen
  \bibfield  {author} {\bibinfo {author} {\bibfnamefont {D.}~\bibnamefont
  {Oshima}}, \bibinfo {author} {\bibfnamefont {K.}~\bibnamefont {Taguchi}},\
  and\ \bibinfo {author} {\bibfnamefont {Y.}~\bibnamefont {Tanaka}},\
  }\bibfield  {title} {\bibinfo {title} {Tunneling conductance in
  two-dimensional junctions between a normal metal and a ferromagnetic rashba
  metal},\ }\href {https://journals.jps.jp/doi/10.7566/JPSJ.87.034710}
  {\bibfield  {journal} {\bibinfo  {journal} {J. Phys. Soc. Japan}\ }\textbf
  {\bibinfo {volume} {87}},\ \bibinfo {pages} {034710} (\bibinfo {year}
  {2018})}\BibitemShut {NoStop}%
\bibitem [{\citenamefont {Oshima}\ \emph {et~al.}(2019)\citenamefont {Oshima},
  \citenamefont {Taguchi},\ and\ \citenamefont
  {Tanaka}}]{oshima2019unconventional}%
  \BibitemOpen
  \bibfield  {author} {\bibinfo {author} {\bibfnamefont {D.}~\bibnamefont
  {Oshima}}, \bibinfo {author} {\bibfnamefont {K.}~\bibnamefont {Taguchi}},\
  and\ \bibinfo {author} {\bibfnamefont {Y.}~\bibnamefont {Tanaka}},\
  }\bibfield  {title} {\bibinfo {title} {Unconventional gate voltage dependence
  of the charge conductance caused by spin-splitting fermi surface by
  rashba-type spin-orbit coupling},\ }\href
  {https://www.sciencedirect.com/science/article/pii/S138694771930387X}
  {\bibfield  {journal} {\bibinfo  {journal} {Physica E: Low Dimens. Syst.
  Nanostruct.}\ }\textbf {\bibinfo {volume} {114}},\ \bibinfo {pages} {113615}
  (\bibinfo {year} {2019})}\BibitemShut {NoStop}%
\bibitem [{\citenamefont {H{\"u}fner}(2013)}]{hufner2013photoelectron}%
  \BibitemOpen
  \bibfield  {author} {\bibinfo {author} {\bibfnamefont {S.}~\bibnamefont
  {H{\"u}fner}},\ }\href@noop {} {\emph {\bibinfo {title} {Photoelectron
  spectroscopy: principles and applications}}}\ (\bibinfo  {publisher}
  {Springer Science \& Business Media},\ \bibinfo {year} {2013})\BibitemShut
  {NoStop}%
\bibitem [{\citenamefont {Lv}\ \emph {et~al.}(2019)\citenamefont {Lv},
  \citenamefont {Qian},\ and\ \citenamefont {Ding}}]{lv2019angle}%
  \BibitemOpen
  \bibfield  {author} {\bibinfo {author} {\bibfnamefont {B.}~\bibnamefont
  {Lv}}, \bibinfo {author} {\bibfnamefont {T.}~\bibnamefont {Qian}},\ and\
  \bibinfo {author} {\bibfnamefont {H.}~\bibnamefont {Ding}},\ }\bibfield
  {title} {\bibinfo {title} {Angle-resolved photoemission spectroscopy and its
  application to topological materials},\ }\href
  {https://www.nature.com/articles/s42254-019-0088-5} {\bibfield  {journal}
  {\bibinfo  {journal} {Nat. Rev. Phys.}\ }\textbf {\bibinfo {volume} {1}},\
  \bibinfo {pages} {609} (\bibinfo {year} {2019})}\BibitemShut {NoStop}%
\bibitem [{\citenamefont {Yu}\ \emph {et~al.}(2020)\citenamefont {Yu},
  \citenamefont {Matt}, \citenamefont {Bisti}, \citenamefont {Wang},
  \citenamefont {Schmitt}, \citenamefont {Chang}, \citenamefont {Eisaki},
  \citenamefont {Feng},\ and\ \citenamefont {Strocov}}]{yu2020relevance}%
  \BibitemOpen
  \bibfield  {author} {\bibinfo {author} {\bibfnamefont {T.}~\bibnamefont
  {Yu}}, \bibinfo {author} {\bibfnamefont {C.~E.}\ \bibnamefont {Matt}},
  \bibinfo {author} {\bibfnamefont {F.}~\bibnamefont {Bisti}}, \bibinfo
  {author} {\bibfnamefont {X.}~\bibnamefont {Wang}}, \bibinfo {author}
  {\bibfnamefont {T.}~\bibnamefont {Schmitt}}, \bibinfo {author} {\bibfnamefont
  {J.}~\bibnamefont {Chang}}, \bibinfo {author} {\bibfnamefont
  {H.}~\bibnamefont {Eisaki}}, \bibinfo {author} {\bibfnamefont
  {D.}~\bibnamefont {Feng}},\ and\ \bibinfo {author} {\bibfnamefont {V.~N.}\
  \bibnamefont {Strocov}},\ }\bibfield  {title} {\bibinfo {title} {The
  relevance of arpes to high-t c superconductivity in cuprates},\ }\href
  {https://www.nature.com/articles/s41535-020-0251-3} {\bibfield  {journal}
  {\bibinfo  {journal} {npj Quantum Mater.}\ }\textbf {\bibinfo {volume} {5}},\
  \bibinfo {pages} {46} (\bibinfo {year} {2020})}\BibitemShut {NoStop}%
\bibitem [{\citenamefont {Shimojima}\ \emph {et~al.}(2015)\citenamefont
  {Shimojima}, \citenamefont {Okazaki},\ and\ \citenamefont
  {Shin}}]{doi:10.7566/JPSJ.84.072001}%
  \BibitemOpen
  \bibfield  {author} {\bibinfo {author} {\bibfnamefont {T.}~\bibnamefont
  {Shimojima}}, \bibinfo {author} {\bibfnamefont {K.}~\bibnamefont {Okazaki}},\
  and\ \bibinfo {author} {\bibfnamefont {S.}~\bibnamefont {Shin}},\ }\bibfield
  {title} {\bibinfo {title} {Low-temperature and high-energy-resolution laser
  photoemission spectroscopy},\ }\href {https://doi.org/10.7566/JPSJ.84.072001}
  {\bibfield  {journal} {\bibinfo  {journal} {J. Phys. Soc. Japan}\ }\textbf
  {\bibinfo {volume} {84}},\ \bibinfo {pages} {072001} (\bibinfo {year}
  {2015})}\BibitemShut {NoStop}%
\bibitem [{\citenamefont {Sobota}\ \emph {et~al.}(2021)\citenamefont {Sobota},
  \citenamefont {He},\ and\ \citenamefont {Shen}}]{RevModPhys.93.025006}%
  \BibitemOpen
  \bibfield  {author} {\bibinfo {author} {\bibfnamefont {J.~A.}\ \bibnamefont
  {Sobota}}, \bibinfo {author} {\bibfnamefont {Y.}~\bibnamefont {He}},\ and\
  \bibinfo {author} {\bibfnamefont {Z.-X.}\ \bibnamefont {Shen}},\ }\bibfield
  {title} {\bibinfo {title} {Angle-resolved photoemission studies of quantum
  materials},\ }\href {https://doi.org/10.1103/RevModPhys.93.025006} {\bibfield
   {journal} {\bibinfo  {journal} {Rev. Mod. Phys.}\ }\textbf {\bibinfo
  {volume} {93}},\ \bibinfo {pages} {025006} (\bibinfo {year}
  {2021})}\BibitemShut {NoStop}%
\bibitem [{\citenamefont {Kornich}\ and\ \citenamefont
  {Trauzettel}(2022)}]{PhysRevResearch.4.L022018}%
  \BibitemOpen
  \bibfield  {author} {\bibinfo {author} {\bibfnamefont {V.}~\bibnamefont
  {Kornich}}\ and\ \bibinfo {author} {\bibfnamefont {B.}~\bibnamefont
  {Trauzettel}},\ }\bibfield  {title} {\bibinfo {title} {Signature of
  $\mathcal{P}\mathcal{T}$-symmetric non-hermitian superconductivity in
  angle-resolved photoelectron fluctuation spectroscopy},\ }\href
  {https://doi.org/10.1103/PhysRevResearch.4.L022018} {\bibfield  {journal}
  {\bibinfo  {journal} {Phys. Rev. Research}\ }\textbf {\bibinfo {volume}
  {4}},\ \bibinfo {pages} {L022018} (\bibinfo {year} {2022})}\BibitemShut
  {NoStop}%
\bibitem [{\citenamefont {Mahan}(2013)}]{mahan2013many}%
  \BibitemOpen
  \bibfield  {author} {\bibinfo {author} {\bibfnamefont {G.~D.}\ \bibnamefont
  {Mahan}},\ }\href@noop {} {\emph {\bibinfo {title} {Many-particle physics}}}\
  (\bibinfo  {publisher} {Springer Science \& Business Media},\ \bibinfo {year}
  {2013})\BibitemShut {NoStop}%
\bibitem [{\citenamefont {Zagoskin}(2014)}]{zagoskin}%
  \BibitemOpen
  \bibfield  {author} {\bibinfo {author} {\bibfnamefont {A.}~\bibnamefont
  {Zagoskin}},\ }\href@noop {} {\emph {\bibinfo {title} {Quantum Theory of
  Many-Body Systems: Techniques and Applications}}}\ (\bibinfo  {publisher}
  {Springer},\ \bibinfo {year} {2014})\BibitemShut {NoStop}%
\bibitem [{\citenamefont {Kj{\ae}rgaard}\ \emph {et~al.}(2016)\citenamefont
  {Kj{\ae}rgaard}, \citenamefont {Nichele}, \citenamefont {Suominen},
  \citenamefont {Nowak}, \citenamefont {Wimmer}, \citenamefont {Akhmerov},
  \citenamefont {Folk}, \citenamefont {Flensberg}, \citenamefont {Shabani},
  \citenamefont {Palmstr{\o}m},\ and\ \citenamefont {Marcus}}]{Kjaergaard}%
  \BibitemOpen
  \bibfield  {author} {\bibinfo {author} {\bibfnamefont {M.}~\bibnamefont
  {Kj{\ae}rgaard}}, \bibinfo {author} {\bibfnamefont {F.}~\bibnamefont
  {Nichele}}, \bibinfo {author} {\bibfnamefont {H.~J.}\ \bibnamefont
  {Suominen}}, \bibinfo {author} {\bibfnamefont {M.~P.}\ \bibnamefont {Nowak}},
  \bibinfo {author} {\bibfnamefont {M.}~\bibnamefont {Wimmer}}, \bibinfo
  {author} {\bibfnamefont {A.~R.}\ \bibnamefont {Akhmerov}}, \bibinfo {author}
  {\bibfnamefont {J.~A.}\ \bibnamefont {Folk}}, \bibinfo {author}
  {\bibfnamefont {K.}~\bibnamefont {Flensberg}}, \bibinfo {author}
  {\bibfnamefont {J.}~\bibnamefont {Shabani}}, \bibinfo {author} {\bibfnamefont
  {C.~J.}\ \bibnamefont {Palmstr{\o}m}},\ and\ \bibinfo {author} {\bibfnamefont
  {C.~M.}\ \bibnamefont {Marcus}},\ }\bibfield  {title} {\bibinfo {title}
  {Quantized conductance doubling and hard gap in a two-dimensional
  semiconductor-superconductor heterostructure},\ }\href
  {https://www.nature.com/articles/ncomms12841} {\bibfield  {journal} {\bibinfo
   {journal} {Nat. Commun.}\ }\textbf {\bibinfo {volume} {7}},\ \bibinfo
  {pages} {12841} (\bibinfo {year} {2016})}\BibitemShut {NoStop}%
\bibitem [{\citenamefont {Shabani}\ \emph {et~al.}(2016)\citenamefont
  {Shabani}, \citenamefont {Kjaergaard}, \citenamefont {Suominen},
  \citenamefont {Kim}, \citenamefont {Nichele}, \citenamefont {Pakrouski},
  \citenamefont {Stankevic}, \citenamefont {Lutchyn}, \citenamefont
  {Krogstrup}, \citenamefont {Feidenhans'l}, \citenamefont {Kraemer},
  \citenamefont {Nayak}, \citenamefont {Troyer}, \citenamefont {Marcus},\ and\
  \citenamefont {Palmstr\o{}m}}]{PhysRevB.93.155402}%
  \BibitemOpen
  \bibfield  {author} {\bibinfo {author} {\bibfnamefont {J.}~\bibnamefont
  {Shabani}}, \bibinfo {author} {\bibfnamefont {M.}~\bibnamefont {Kjaergaard}},
  \bibinfo {author} {\bibfnamefont {H.~J.}\ \bibnamefont {Suominen}}, \bibinfo
  {author} {\bibfnamefont {Y.}~\bibnamefont {Kim}}, \bibinfo {author}
  {\bibfnamefont {F.}~\bibnamefont {Nichele}}, \bibinfo {author} {\bibfnamefont
  {K.}~\bibnamefont {Pakrouski}}, \bibinfo {author} {\bibfnamefont
  {T.}~\bibnamefont {Stankevic}}, \bibinfo {author} {\bibfnamefont {R.~M.}\
  \bibnamefont {Lutchyn}}, \bibinfo {author} {\bibfnamefont {P.}~\bibnamefont
  {Krogstrup}}, \bibinfo {author} {\bibfnamefont {R.}~\bibnamefont
  {Feidenhans'l}}, \bibinfo {author} {\bibfnamefont {S.}~\bibnamefont
  {Kraemer}}, \bibinfo {author} {\bibfnamefont {C.}~\bibnamefont {Nayak}},
  \bibinfo {author} {\bibfnamefont {M.}~\bibnamefont {Troyer}}, \bibinfo
  {author} {\bibfnamefont {C.~M.}\ \bibnamefont {Marcus}},\ and\ \bibinfo
  {author} {\bibfnamefont {C.~J.}\ \bibnamefont {Palmstr\o{}m}},\ }\bibfield
  {title} {\bibinfo {title} {Two-dimensional epitaxial
  superconductor-semiconductor heterostructures: A platform for topological
  superconducting networks},\ }\href
  {https://doi.org/10.1103/PhysRevB.93.155402} {\bibfield  {journal} {\bibinfo
  {journal} {Phys. Rev. B}\ }\textbf {\bibinfo {volume} {93}},\ \bibinfo
  {pages} {155402} (\bibinfo {year} {2016})}\BibitemShut {NoStop}%
\bibitem [{\citenamefont {Suominen}\ \emph {et~al.}(2017)\citenamefont
  {Suominen}, \citenamefont {Kjaergaard}, \citenamefont {Hamilton},
  \citenamefont {Shabani}, \citenamefont {Palmstr\o{}m}, \citenamefont
  {Marcus},\ and\ \citenamefont {Nichele}}]{Suominen17}%
  \BibitemOpen
  \bibfield  {author} {\bibinfo {author} {\bibfnamefont {H.~J.}\ \bibnamefont
  {Suominen}}, \bibinfo {author} {\bibfnamefont {M.}~\bibnamefont
  {Kjaergaard}}, \bibinfo {author} {\bibfnamefont {A.~R.}\ \bibnamefont
  {Hamilton}}, \bibinfo {author} {\bibfnamefont {J.}~\bibnamefont {Shabani}},
  \bibinfo {author} {\bibfnamefont {C.~J.}\ \bibnamefont {Palmstr\o{}m}},
  \bibinfo {author} {\bibfnamefont {C.~M.}\ \bibnamefont {Marcus}},\ and\
  \bibinfo {author} {\bibfnamefont {F.}~\bibnamefont {Nichele}},\ }\bibfield
  {title} {\bibinfo {title} {Zero-energy modes from coalescing {A}ndreev states
  in a two-dimensional semiconductor-superconductor hybrid platform},\ }\href
  {https://link.aps.org/doi/10.1103/PhysRevLett.119.176805} {\bibfield
  {journal} {\bibinfo  {journal} {Phys. Rev. Lett.}\ }\textbf {\bibinfo
  {volume} {119}},\ \bibinfo {pages} {176805} (\bibinfo {year}
  {2017})}\BibitemShut {NoStop}%
\bibitem [{\citenamefont {B{\o}ttcher}\ \emph {et~al.}(2018)\citenamefont
  {B{\o}ttcher}, \citenamefont {Nichele}, \citenamefont {Kjaergaard},
  \citenamefont {Suominen}, \citenamefont {Shabani}, \citenamefont
  {Palmstr{\o}m},\ and\ \citenamefont {Marcus}}]{bottcher2018superconducting}%
  \BibitemOpen
  \bibfield  {author} {\bibinfo {author} {\bibfnamefont {C.}~\bibnamefont
  {B{\o}ttcher}}, \bibinfo {author} {\bibfnamefont {F.}~\bibnamefont
  {Nichele}}, \bibinfo {author} {\bibfnamefont {M.}~\bibnamefont {Kjaergaard}},
  \bibinfo {author} {\bibfnamefont {H.}~\bibnamefont {Suominen}}, \bibinfo
  {author} {\bibfnamefont {J.}~\bibnamefont {Shabani}}, \bibinfo {author}
  {\bibfnamefont {C.}~\bibnamefont {Palmstr{\o}m}},\ and\ \bibinfo {author}
  {\bibfnamefont {C.}~\bibnamefont {Marcus}},\ }\bibfield  {title} {\bibinfo
  {title} {Superconducting, insulating and anomalous metallic regimes in a
  gated two-dimensional semiconductor--superconductor array},\ }\href
  {https://www.nature.com/articles/s41567-018-0259-9} {\bibfield  {journal}
  {\bibinfo  {journal} {Nat. Phys.}\ }\textbf {\bibinfo {volume} {14}},\
  \bibinfo {pages} {1138} (\bibinfo {year} {2018})}\BibitemShut {NoStop}%
\bibitem [{\citenamefont {Fornieri}\ \emph {et~al.}(2019)\citenamefont
  {Fornieri}, \citenamefont {Whiticar}, \citenamefont {Setiawan}, \citenamefont
  {Portol{\'e}s}, \citenamefont {Drachmann}, \citenamefont {Keselman},
  \citenamefont {Gronin}, \citenamefont {Thomas}, \citenamefont {Wang},
  \citenamefont {Kallaher} \emph {et~al.}}]{fornieri2019evidence}%
  \BibitemOpen
  \bibfield  {author} {\bibinfo {author} {\bibfnamefont {A.}~\bibnamefont
  {Fornieri}}, \bibinfo {author} {\bibfnamefont {A.~M.}\ \bibnamefont
  {Whiticar}}, \bibinfo {author} {\bibfnamefont {F.}~\bibnamefont {Setiawan}},
  \bibinfo {author} {\bibfnamefont {E.}~\bibnamefont {Portol{\'e}s}}, \bibinfo
  {author} {\bibfnamefont {A.~C.}\ \bibnamefont {Drachmann}}, \bibinfo {author}
  {\bibfnamefont {A.}~\bibnamefont {Keselman}}, \bibinfo {author}
  {\bibfnamefont {S.}~\bibnamefont {Gronin}}, \bibinfo {author} {\bibfnamefont
  {C.}~\bibnamefont {Thomas}}, \bibinfo {author} {\bibfnamefont
  {T.}~\bibnamefont {Wang}}, \bibinfo {author} {\bibfnamefont {R.}~\bibnamefont
  {Kallaher}}, \emph {et~al.},\ }\bibfield  {title} {\bibinfo {title} {Evidence
  of topological superconductivity in planar josephson junctions},\ }\href
  {https://www.nature.com/articles/s41586-019-1068-8} {\bibfield  {journal}
  {\bibinfo  {journal} {Nature}\ }\textbf {\bibinfo {volume} {569}},\ \bibinfo
  {pages} {89} (\bibinfo {year} {2019})}\BibitemShut {NoStop}%
\bibitem [{\citenamefont {O’Connell~Yuan}\ \emph {et~al.}(2021)\citenamefont
  {O’Connell~Yuan}, \citenamefont {Wickramasinghe}, \citenamefont
  {Strickland}, \citenamefont {Dartiailh}, \citenamefont {Sardashti},
  \citenamefont {Hatefipour},\ and\ \citenamefont {Shabani}}]{o2021epitaxial}%
  \BibitemOpen
  \bibfield  {author} {\bibinfo {author} {\bibfnamefont {J.}~\bibnamefont
  {O’Connell~Yuan}}, \bibinfo {author} {\bibfnamefont {K.~S.}\ \bibnamefont
  {Wickramasinghe}}, \bibinfo {author} {\bibfnamefont {W.~M.}\ \bibnamefont
  {Strickland}}, \bibinfo {author} {\bibfnamefont {M.~C.}\ \bibnamefont
  {Dartiailh}}, \bibinfo {author} {\bibfnamefont {K.}~\bibnamefont
  {Sardashti}}, \bibinfo {author} {\bibfnamefont {M.}~\bibnamefont
  {Hatefipour}},\ and\ \bibinfo {author} {\bibfnamefont {J.}~\bibnamefont
  {Shabani}},\ }\bibfield  {title} {\bibinfo {title} {Epitaxial
  superconductor-semiconductor two-dimensional systems for superconducting
  quantum circuits},\ }\href {https://avs.scitation.org/doi/10.1116/6.0000918}
  {\bibfield  {journal} {\bibinfo  {journal} {J. Vac. Sci. Technol. A: Vac.
  Surf.}\ }\textbf {\bibinfo {volume} {39}},\ \bibinfo {pages} {033407}
  (\bibinfo {year} {2021})}\BibitemShut {NoStop}%
\bibitem [{\citenamefont {Gazibegovic}\ \emph {et~al.}(2019)\citenamefont
  {Gazibegovic}, \citenamefont {Badawy}, \citenamefont {Buckers}, \citenamefont
  {Leubner}, \citenamefont {Shen}, \citenamefont {de~Vries}, \citenamefont
  {Koelling}, \citenamefont {Kouwenhoven}, \citenamefont {Verheijen},\ and\
  \citenamefont {Bakkers}}]{gazibegovic2019bottom}%
  \BibitemOpen
  \bibfield  {author} {\bibinfo {author} {\bibfnamefont {S.}~\bibnamefont
  {Gazibegovic}}, \bibinfo {author} {\bibfnamefont {G.}~\bibnamefont {Badawy}},
  \bibinfo {author} {\bibfnamefont {T.~L.}\ \bibnamefont {Buckers}}, \bibinfo
  {author} {\bibfnamefont {P.}~\bibnamefont {Leubner}}, \bibinfo {author}
  {\bibfnamefont {J.}~\bibnamefont {Shen}}, \bibinfo {author} {\bibfnamefont
  {F.~K.}\ \bibnamefont {de~Vries}}, \bibinfo {author} {\bibfnamefont
  {S.}~\bibnamefont {Koelling}}, \bibinfo {author} {\bibfnamefont {L.~P.}\
  \bibnamefont {Kouwenhoven}}, \bibinfo {author} {\bibfnamefont {M.~A.}\
  \bibnamefont {Verheijen}},\ and\ \bibinfo {author} {\bibfnamefont {E.~P.}\
  \bibnamefont {Bakkers}},\ }\bibfield  {title} {\bibinfo {title} {Bottom-up
  grown 2d insb nanostructures},\ }\href
  {https://onlinelibrary.wiley.com/doi/full/10.1002/adma.201808181} {\bibfield
  {journal} {\bibinfo  {journal} {Adv. Mater.}\ }\textbf {\bibinfo {volume}
  {31}},\ \bibinfo {pages} {1808181} (\bibinfo {year} {2019})}\BibitemShut
  {NoStop}%
\bibitem [{\citenamefont {Ke}\ \emph {et~al.}(2019)\citenamefont {Ke},
  \citenamefont {Moehle}, \citenamefont {de~Vries}, \citenamefont {Thomas},
  \citenamefont {Metti}, \citenamefont {Guinn}, \citenamefont {Kallaher},
  \citenamefont {Lodari}, \citenamefont {Scappucci}, \citenamefont {Wang} \emph
  {et~al.}}]{ke2019ballistic}%
  \BibitemOpen
  \bibfield  {author} {\bibinfo {author} {\bibfnamefont {C.~T.}\ \bibnamefont
  {Ke}}, \bibinfo {author} {\bibfnamefont {C.~M.}\ \bibnamefont {Moehle}},
  \bibinfo {author} {\bibfnamefont {F.~K.}\ \bibnamefont {de~Vries}}, \bibinfo
  {author} {\bibfnamefont {C.}~\bibnamefont {Thomas}}, \bibinfo {author}
  {\bibfnamefont {S.}~\bibnamefont {Metti}}, \bibinfo {author} {\bibfnamefont
  {C.~R.}\ \bibnamefont {Guinn}}, \bibinfo {author} {\bibfnamefont
  {R.}~\bibnamefont {Kallaher}}, \bibinfo {author} {\bibfnamefont
  {M.}~\bibnamefont {Lodari}}, \bibinfo {author} {\bibfnamefont
  {G.}~\bibnamefont {Scappucci}}, \bibinfo {author} {\bibfnamefont
  {T.}~\bibnamefont {Wang}}, \emph {et~al.},\ }\bibfield  {title} {\bibinfo
  {title} {Ballistic superconductivity and tunable $\pi$--junctions in insb
  quantum wells},\ }\href {https://www.nature.com/articles/s41467-019-11742-4}
  {\bibfield  {journal} {\bibinfo  {journal} {Nat. Commun.}\ }\textbf {\bibinfo
  {volume} {10}},\ \bibinfo {pages} {3764} (\bibinfo {year}
  {2019})}\BibitemShut {NoStop}%
\bibitem [{\citenamefont {Xue}\ \emph {et~al.}(2019)\citenamefont {Xue},
  \citenamefont {Chen}, \citenamefont {Pan}, \citenamefont {Wang},
  \citenamefont {Zhao}, \citenamefont {Huang},\ and\ \citenamefont
  {Xu}}]{xue2019gate}%
  \BibitemOpen
  \bibfield  {author} {\bibinfo {author} {\bibfnamefont {J.}~\bibnamefont
  {Xue}}, \bibinfo {author} {\bibfnamefont {Y.}~\bibnamefont {Chen}}, \bibinfo
  {author} {\bibfnamefont {D.}~\bibnamefont {Pan}}, \bibinfo {author}
  {\bibfnamefont {J.-Y.}\ \bibnamefont {Wang}}, \bibinfo {author}
  {\bibfnamefont {J.}~\bibnamefont {Zhao}}, \bibinfo {author} {\bibfnamefont
  {S.}~\bibnamefont {Huang}},\ and\ \bibinfo {author} {\bibfnamefont
  {H.}~\bibnamefont {Xu}},\ }\bibfield  {title} {\bibinfo {title} {Gate defined
  quantum dot realized in a single crystalline insb nanosheet},\ }\href
  {https://aip.scitation.org/doi/10.1063/1.5064368} {\bibfield  {journal}
  {\bibinfo  {journal} {Appl. Phys. Lett.}\ }\textbf {\bibinfo {volume}
  {114}},\ \bibinfo {pages} {023108} (\bibinfo {year} {2019})}\BibitemShut
  {NoStop}%
\bibitem [{\citenamefont {Lei}\ \emph {et~al.}(2019)\citenamefont {Lei},
  \citenamefont {Lehner}, \citenamefont {Cheah}, \citenamefont {Karalic},
  \citenamefont {Mittag}, \citenamefont {Alt}, \citenamefont {Scharnetzky},
  \citenamefont {Wegscheider}, \citenamefont {Ihn},\ and\ \citenamefont
  {Ensslin}}]{lei2019quantum}%
  \BibitemOpen
  \bibfield  {author} {\bibinfo {author} {\bibfnamefont {Z.}~\bibnamefont
  {Lei}}, \bibinfo {author} {\bibfnamefont {C.~A.}\ \bibnamefont {Lehner}},
  \bibinfo {author} {\bibfnamefont {E.}~\bibnamefont {Cheah}}, \bibinfo
  {author} {\bibfnamefont {M.}~\bibnamefont {Karalic}}, \bibinfo {author}
  {\bibfnamefont {C.}~\bibnamefont {Mittag}}, \bibinfo {author} {\bibfnamefont
  {L.}~\bibnamefont {Alt}}, \bibinfo {author} {\bibfnamefont {J.}~\bibnamefont
  {Scharnetzky}}, \bibinfo {author} {\bibfnamefont {W.}~\bibnamefont
  {Wegscheider}}, \bibinfo {author} {\bibfnamefont {T.}~\bibnamefont {Ihn}},\
  and\ \bibinfo {author} {\bibfnamefont {K.}~\bibnamefont {Ensslin}},\
  }\bibfield  {title} {\bibinfo {title} {Quantum transport in high-quality
  shallow insb quantum wells},\ }\href
  {https://aip.scitation.org/doi/10.1063/1.5098294} {\bibfield  {journal}
  {\bibinfo  {journal} {Appl. Phys. Lett.}\ }\textbf {\bibinfo {volume}
  {115}},\ \bibinfo {pages} {012101} (\bibinfo {year} {2019})}\BibitemShut
  {NoStop}%
\bibitem [{\citenamefont {Chen}\ \emph
  {et~al.}(2021{\natexlab{b}})\citenamefont {Chen}, \citenamefont {Huang},
  \citenamefont {Pan}, \citenamefont {Xue}, \citenamefont {Zhang},
  \citenamefont {Zhao},\ and\ \citenamefont {Xu}}]{chen2021strong}%
  \BibitemOpen
  \bibfield  {author} {\bibinfo {author} {\bibfnamefont {Y.}~\bibnamefont
  {Chen}}, \bibinfo {author} {\bibfnamefont {S.}~\bibnamefont {Huang}},
  \bibinfo {author} {\bibfnamefont {D.}~\bibnamefont {Pan}}, \bibinfo {author}
  {\bibfnamefont {J.}~\bibnamefont {Xue}}, \bibinfo {author} {\bibfnamefont
  {L.}~\bibnamefont {Zhang}}, \bibinfo {author} {\bibfnamefont
  {J.}~\bibnamefont {Zhao}},\ and\ \bibinfo {author} {\bibfnamefont
  {H.}~\bibnamefont {Xu}},\ }\bibfield  {title} {\bibinfo {title} {Strong and
  tunable spin--orbit interaction in a single crystalline insb nanosheet},\
  }\href {https://www.nature.com/articles/s41699-020-00184-y} {\bibfield
  {journal} {\bibinfo  {journal} {npj 2D Mater. Appl.}\ }\textbf {\bibinfo
  {volume} {5}},\ \bibinfo {pages} {3} (\bibinfo {year}
  {2021}{\natexlab{b}})}\BibitemShut {NoStop}%
\bibitem [{\citenamefont {Katmis}\ \emph {et~al.}(2016)\citenamefont {Katmis},
  \citenamefont {Lauter}, \citenamefont {Nogueira}, \citenamefont {Assaf},
  \citenamefont {Jamer}, \citenamefont {Wei}, \citenamefont {Satpati},
  \citenamefont {Freeland}, \citenamefont {Eremin}, \citenamefont {Heiman}
  \emph {et~al.}}]{katmis2016high}%
  \BibitemOpen
  \bibfield  {author} {\bibinfo {author} {\bibfnamefont {F.}~\bibnamefont
  {Katmis}}, \bibinfo {author} {\bibfnamefont {V.}~\bibnamefont {Lauter}},
  \bibinfo {author} {\bibfnamefont {F.~S.}\ \bibnamefont {Nogueira}}, \bibinfo
  {author} {\bibfnamefont {B.~A.}\ \bibnamefont {Assaf}}, \bibinfo {author}
  {\bibfnamefont {M.~E.}\ \bibnamefont {Jamer}}, \bibinfo {author}
  {\bibfnamefont {P.}~\bibnamefont {Wei}}, \bibinfo {author} {\bibfnamefont
  {B.}~\bibnamefont {Satpati}}, \bibinfo {author} {\bibfnamefont {J.~W.}\
  \bibnamefont {Freeland}}, \bibinfo {author} {\bibfnamefont {I.}~\bibnamefont
  {Eremin}}, \bibinfo {author} {\bibfnamefont {D.}~\bibnamefont {Heiman}},
  \emph {et~al.},\ }\bibfield  {title} {\bibinfo {title} {A high-temperature
  ferromagnetic topological insulating phase by proximity coupling},\ }\href
  {https://www.nature.com/articles/nature17635} {\bibfield  {journal} {\bibinfo
   {journal} {Nature}\ }\textbf {\bibinfo {volume} {533}},\ \bibinfo {pages}
  {513} (\bibinfo {year} {2016})}\BibitemShut {NoStop}%
\bibitem [{\citenamefont {Liu}\ \emph {et~al.}(2019)\citenamefont {Liu},
  \citenamefont {Vaitiekenas}, \citenamefont {Mart{\'\i}-S{\'a}nchez},
  \citenamefont {Koch}, \citenamefont {Hart}, \citenamefont {Cui},
  \citenamefont {Kanne}, \citenamefont {Khan}, \citenamefont {Tanta},
  \citenamefont {Upadhyay}, \citenamefont {Cachaza}, \citenamefont {Marcus},
  \citenamefont {Arbiol}, \citenamefont {Moler},\ and\ \citenamefont
  {Krogstrup}}]{liu2019semiconductor}%
  \BibitemOpen
  \bibfield  {author} {\bibinfo {author} {\bibfnamefont {Y.}~\bibnamefont
  {Liu}}, \bibinfo {author} {\bibfnamefont {S.}~\bibnamefont {Vaitiekenas}},
  \bibinfo {author} {\bibfnamefont {S.}~\bibnamefont {Mart{\'\i}-S{\'a}nchez}},
  \bibinfo {author} {\bibfnamefont {C.}~\bibnamefont {Koch}}, \bibinfo {author}
  {\bibfnamefont {S.}~\bibnamefont {Hart}}, \bibinfo {author} {\bibfnamefont
  {Z.}~\bibnamefont {Cui}}, \bibinfo {author} {\bibfnamefont {T.}~\bibnamefont
  {Kanne}}, \bibinfo {author} {\bibfnamefont {S.~A.}\ \bibnamefont {Khan}},
  \bibinfo {author} {\bibfnamefont {R.}~\bibnamefont {Tanta}}, \bibinfo
  {author} {\bibfnamefont {S.}~\bibnamefont {Upadhyay}}, \bibinfo {author}
  {\bibfnamefont {M.~E.}\ \bibnamefont {Cachaza}}, \bibinfo {author}
  {\bibfnamefont {C.~M.}\ \bibnamefont {Marcus}}, \bibinfo {author}
  {\bibfnamefont {J.}~\bibnamefont {Arbiol}}, \bibinfo {author} {\bibfnamefont
  {K.~A.}\ \bibnamefont {Moler}},\ and\ \bibinfo {author} {\bibfnamefont
  {P.}~\bibnamefont {Krogstrup}},\ }\bibfield  {title} {\bibinfo {title}
  {Semiconductor--ferromagnetic insulator--superconductor nanowires: {S}tray
  field and exchange field},\ }\href
  {https://pubs.acs.org/doi/full/10.1021/acs.nanolett.9b04187} {\bibfield
  {journal} {\bibinfo  {journal} {Nano Lett.}\ }\textbf {\bibinfo {volume}
  {20}},\ \bibinfo {pages} {456} (\bibinfo {year} {2019})}\BibitemShut
  {NoStop}%
\bibitem [{\citenamefont {Yang}\ \emph {et~al.}(2020)\citenamefont {Yang},
  \citenamefont {Heischmidt}, \citenamefont {Gazibegovic}, \citenamefont
  {Badawy}, \citenamefont {Car}, \citenamefont {Crowell}, \citenamefont
  {Bakkers},\ and\ \citenamefont {Pribiag}}]{yang2020spin}%
  \BibitemOpen
  \bibfield  {author} {\bibinfo {author} {\bibfnamefont {Z.}~\bibnamefont
  {Yang}}, \bibinfo {author} {\bibfnamefont {B.}~\bibnamefont {Heischmidt}},
  \bibinfo {author} {\bibfnamefont {S.}~\bibnamefont {Gazibegovic}}, \bibinfo
  {author} {\bibfnamefont {G.}~\bibnamefont {Badawy}}, \bibinfo {author}
  {\bibfnamefont {D.}~\bibnamefont {Car}}, \bibinfo {author} {\bibfnamefont
  {P.~A.}\ \bibnamefont {Crowell}}, \bibinfo {author} {\bibfnamefont {E.~P.}\
  \bibnamefont {Bakkers}},\ and\ \bibinfo {author} {\bibfnamefont {V.~S.}\
  \bibnamefont {Pribiag}},\ }\bibfield  {title} {\bibinfo {title} {Spin
  transport in ferromagnet-insb nanowire quantum devices},\ }\href
  {https://pubs.acs.org/doi/full/10.1021/acs.nanolett.9b05331} {\bibfield
  {journal} {\bibinfo  {journal} {Nano Lett.}\ }\textbf {\bibinfo {volume}
  {20}},\ \bibinfo {pages} {3232} (\bibinfo {year} {2020})}\BibitemShut
  {NoStop}%
\bibitem [{\citenamefont {Vaitiek{\.e}nas}\ \emph {et~al.}(2021)\citenamefont
  {Vaitiek{\.e}nas}, \citenamefont {Liu}, \citenamefont {Krogstrup},\ and\
  \citenamefont {Marcus}}]{vaitiekenas2021zero}%
  \BibitemOpen
  \bibfield  {author} {\bibinfo {author} {\bibfnamefont {S.}~\bibnamefont
  {Vaitiek{\.e}nas}}, \bibinfo {author} {\bibfnamefont {Y.}~\bibnamefont
  {Liu}}, \bibinfo {author} {\bibfnamefont {P.}~\bibnamefont {Krogstrup}},\
  and\ \bibinfo {author} {\bibfnamefont {C.}~\bibnamefont {Marcus}},\
  }\bibfield  {title} {\bibinfo {title} {Zero-bias peaks at zero magnetic field
  in ferromagnetic hybrid nanowires},\ }\href
  {https://www.nature.com/articles/s41567-020-1017-3} {\bibfield  {journal}
  {\bibinfo  {journal} {Nat. Phys.}\ }\textbf {\bibinfo {volume} {17}},\
  \bibinfo {pages} {43} (\bibinfo {year} {2021})}\BibitemShut {NoStop}%
\bibitem [{\citenamefont {Escribano}\ \emph {et~al.}(2022)\citenamefont
  {Escribano}, \citenamefont {Maiani}, \citenamefont {Leijnse}, \citenamefont
  {Flensberg}, \citenamefont {Oreg}, \citenamefont {Levy~Yeyati}, \citenamefont
  {Prada},\ and\ \citenamefont {Seoane~Souto}}]{escribano2022semiconductor}%
  \BibitemOpen
  \bibfield  {author} {\bibinfo {author} {\bibfnamefont {S.~D.}\ \bibnamefont
  {Escribano}}, \bibinfo {author} {\bibfnamefont {A.}~\bibnamefont {Maiani}},
  \bibinfo {author} {\bibfnamefont {M.}~\bibnamefont {Leijnse}}, \bibinfo
  {author} {\bibfnamefont {K.}~\bibnamefont {Flensberg}}, \bibinfo {author}
  {\bibfnamefont {Y.}~\bibnamefont {Oreg}}, \bibinfo {author} {\bibfnamefont
  {A.}~\bibnamefont {Levy~Yeyati}}, \bibinfo {author} {\bibfnamefont
  {E.}~\bibnamefont {Prada}},\ and\ \bibinfo {author} {\bibfnamefont
  {R.}~\bibnamefont {Seoane~Souto}},\ }\bibfield  {title} {\bibinfo {title}
  {Semiconductor-ferromagnet-superconductor planar heterostructures for 1d
  topological superconductivity},\ }\href
  {https://www.nature.com/articles/s41535-022-00489-9} {\bibfield  {journal}
  {\bibinfo  {journal} {npj Quantum Mater.}\ }\textbf {\bibinfo {volume} {7}},\
  \bibinfo {pages} {81} (\bibinfo {year} {2022})}\BibitemShut {NoStop}%
\bibitem [{\citenamefont {Vaitiek\ifmmode~\dot{e}\else \.{e}\fi{}nas}\ \emph
  {et~al.}(2022)\citenamefont {Vaitiek\ifmmode~\dot{e}\else \.{e}\fi{}nas},
  \citenamefont {Souto}, \citenamefont {Liu}, \citenamefont {Krogstrup},
  \citenamefont {Flensberg}, \citenamefont {Leijnse},\ and\ \citenamefont
  {Marcus}}]{PhysRevB.105.L041304}%
  \BibitemOpen
  \bibfield  {author} {\bibinfo {author} {\bibfnamefont {S.}~\bibnamefont
  {Vaitiek\ifmmode~\dot{e}\else \.{e}\fi{}nas}}, \bibinfo {author}
  {\bibfnamefont {R.~S.}\ \bibnamefont {Souto}}, \bibinfo {author}
  {\bibfnamefont {Y.}~\bibnamefont {Liu}}, \bibinfo {author} {\bibfnamefont
  {P.}~\bibnamefont {Krogstrup}}, \bibinfo {author} {\bibfnamefont
  {K.}~\bibnamefont {Flensberg}}, \bibinfo {author} {\bibfnamefont
  {M.}~\bibnamefont {Leijnse}},\ and\ \bibinfo {author} {\bibfnamefont {C.~M.}\
  \bibnamefont {Marcus}},\ }\bibfield  {title} {\bibinfo {title} {Evidence for
  spin-polarized bound states in
  semiconductor--superconductor--ferromagnetic-insulator islands},\ }\href
  {https://doi.org/10.1103/PhysRevB.105.L041304} {\bibfield  {journal}
  {\bibinfo  {journal} {Phys. Rev. B}\ }\textbf {\bibinfo {volume} {105}},\
  \bibinfo {pages} {L041304} (\bibinfo {year} {2022})}\BibitemShut {NoStop}%
\bibitem [{\citenamefont {Razmadze}\ \emph {et~al.}(2023)\citenamefont
  {Razmadze}, \citenamefont {Souto}, \citenamefont {Galletti}, \citenamefont
  {Maiani}, \citenamefont {Liu}, \citenamefont {Krogstrup}, \citenamefont
  {Schrade}, \citenamefont {Gyenis}, \citenamefont {Marcus},\ and\
  \citenamefont {Vaitiek\ifmmode~\dot{e}\else
  \.{e}\fi{}nas}}]{razmadze2022supercurrent}%
  \BibitemOpen
  \bibfield  {author} {\bibinfo {author} {\bibfnamefont {D.}~\bibnamefont
  {Razmadze}}, \bibinfo {author} {\bibfnamefont {R.~S.}\ \bibnamefont {Souto}},
  \bibinfo {author} {\bibfnamefont {L.}~\bibnamefont {Galletti}}, \bibinfo
  {author} {\bibfnamefont {A.}~\bibnamefont {Maiani}}, \bibinfo {author}
  {\bibfnamefont {Y.}~\bibnamefont {Liu}}, \bibinfo {author} {\bibfnamefont
  {P.}~\bibnamefont {Krogstrup}}, \bibinfo {author} {\bibfnamefont
  {C.}~\bibnamefont {Schrade}}, \bibinfo {author} {\bibfnamefont
  {A.}~\bibnamefont {Gyenis}}, \bibinfo {author} {\bibfnamefont {C.~M.}\
  \bibnamefont {Marcus}},\ and\ \bibinfo {author} {\bibfnamefont
  {S.}~\bibnamefont {Vaitiek\ifmmode~\dot{e}\else \.{e}\fi{}nas}},\ }\bibfield
  {title} {\bibinfo {title} {Supercurrent reversal in ferromagnetic hybrid
  nanowire josephson junctions},\ }\href
  {https://doi.org/10.1103/PhysRevB.107.L081301} {\bibfield  {journal}
  {\bibinfo  {journal} {Phys. Rev. B}\ }\textbf {\bibinfo {volume} {107}},\
  \bibinfo {pages} {L081301} (\bibinfo {year} {2023})}\BibitemShut {NoStop}%
\bibitem [{\citenamefont {Ryndyk}\ \emph {et~al.}(2009)\citenamefont {Ryndyk},
  \citenamefont {Guti{\'e}rrez}, \citenamefont {Song},\ and\ \citenamefont
  {Cuniberti}}]{Ryndyk2009}%
  \BibitemOpen
  \bibfield  {author} {\bibinfo {author} {\bibfnamefont {D.~A.}\ \bibnamefont
  {Ryndyk}}, \bibinfo {author} {\bibfnamefont {R.}~\bibnamefont
  {Guti{\'e}rrez}}, \bibinfo {author} {\bibfnamefont {B.}~\bibnamefont
  {Song}},\ and\ \bibinfo {author} {\bibfnamefont {G.}~\bibnamefont
  {Cuniberti}},\ }\bibinfo {title} {Green function techniques in the treatment
  of quantum transport at the molecular scale},\ in\ \href
  {https://doi.org/10.1007/978-3-642-02306-4_9} {\emph {\bibinfo {booktitle}
  {Energy Transfer Dynamics in Biomaterial Systems}}},\ \bibinfo {editor}
  {edited by\ \bibinfo {editor} {\bibfnamefont {I.}~\bibnamefont {Burghardt}},
  \bibinfo {editor} {\bibfnamefont {V.}~\bibnamefont {May}}, \bibinfo {editor}
  {\bibfnamefont {D.~A.}\ \bibnamefont {Micha}},\ and\ \bibinfo {editor}
  {\bibfnamefont {E.~R.}\ \bibnamefont {Bittner}}}\ (\bibinfo  {publisher}
  {Springer Berlin Heidelberg},\ \bibinfo {address} {Berlin, Heidelberg},\
  \bibinfo {year} {2009})\ pp.\ \bibinfo {pages} {213--335}\BibitemShut
  {NoStop}%
\end{thebibliography}%
\end{document}